\documentclass[11pt]{article}
\pdfoutput=1
\usepackage[utf8x]{inputenc}	
\usepackage[english]{babel}
\usepackage{amssymb}
\usepackage{amsthm,amsmath}
\usepackage{mathtools}		
\usepackage{dsfont}		
\usepackage{stmaryrd}		
\usepackage{cite}		
\usepackage{youngtab}
\usepackage[svgnames]{xcolor}
\usepackage{enumerate}
\usepackage{setspace}
\usepackage{booktabs}
\usepackage{tikz}
\usetikzlibrary{shapes}
\usetikzlibrary{positioning}
\usetikzlibrary{chains}
\usetikzlibrary{arrows,fit,decorations.pathreplacing, shapes.misc}
\tikzstyle{every picture}+=[remember picture]
\tikzstyle{na} = [baseline=-.5ex]
\usepackage[font={small,it},labelfont=bf]{caption}  
\usepackage{subcaption}
\usepackage[pdftex,breaklinks,colorlinks=true,urlcolor=RoyalBlue,linkcolor=blue,
citecolor=blue]{hyperref}		
\usepackage{paralist}

\catcode`@=11
\renewcommand{\section}{\@startsection{section}{1}{0pt}{\medskipamount}
{\medskipamount}{\Large\bf}}
\numberwithin{equation}{section}
\catcode`@=12

\newcommand{\diff}{\mathrm{d}}
\newcommand{\HS}{\mathrm{HS}}

\newcommand{\PE}{\mathrm{PE}}

\newcommand{\C}{\mathbb{C}}

\newcommand{\NN}{\mathbb{N}}  
\newcommand{\Z}{\mathbb{Z}}
\newcommand{\Mcal}{\mathcal{M}}
\newcommand{\MCoulomb}{\mathcal{M}_C}

\newcommand{\MHiggs}{\mathcal{M}_H}

\newcommand{\clorbit}[1]{\overline{\mathcal{O}}_{#1}}
\newcommand{\height}[1]{\text{ht}(#1)}

\newcommand{\Ncal}{\mathcal{N}}

\newcommand{\gfrak}{\mathfrak{g}}  

\newcommand{\tfrak}{\mathfrak{t}}
  
\newcommand{\Wcal}{\mathcal{W}}

\newcommand{\uo}{{ \mathrm{U}(1)}}

\newcommand{\urm}{{{\rm U}}}

\newcommand{\surm}{{{\rm SU}}}

\newcommand{\sorm}{{{\rm SO}}}
\newcommand{\orm}{{{\rm O}}}

\newcommand{\usprm}{{{\rm USp}}}

\newcommand{\Lie}{\mathrm{Lie}}

\newcommand{\Sym}[1]{\mathrm{Sym}^{#1}}
\newcommand{\Adj}{\mathrm{Adj}}

\newtheorem{myProp}{Proposition}
\newtheorem{myCor}{Corollary}

\allowdisplaybreaks

\begin{document}
\begin{titlepage}
\setcounter{page}{0}
\begin{flushright}
Imperial/TP/18/AH/06\\
UWTHPH-2018-25 
\end{flushright}

\vskip 2cm

\begin{center}

{\Large\bf Discrete quotients of $3$-dimensional $\Ncal=4$ Coulomb branches \\
via the cycle index 
}

\vspace{15mm}

{\large Amihay Hanany${}^{1}$} , \ {\large Marcus Sperling${}^{2}$} 
\\[5mm]
\noindent ${}^1$\emph{Theoretical Physics Group, Imperial College London\\
Prince Consort Road, London, SW7 2AZ, UK}\\
{Email: \texttt{a.hanany@imperial.ac.uk}}
\\[5mm]
\noindent ${}^{2}$\emph{Fakultät für Physik, Universität Wien\\
Boltzmanngasse 5, 1090 Wien, Austria}\\
Email: \texttt{marcus.sperling@univie.ac.at}
\\[5mm]

\vspace{15mm}

\begin{abstract}
The study of Coulomb branches of $3$-dimensional $\Ncal=4$ gauge theories via 
the associated Hilbert series, the so-called monopole formula, has 
been proven useful not only for $3$-dimensional theories, but also for Higgs 
branches of $5$ and $6$-dimensional gauge theories with $8$ supercharges. 
Recently, a conjecture connected different phases of $6$-dimensional Higgs 
branches via \emph{gauging of a discrete global} $S_n$ \emph{symmetry}. 
On the corresponding $3$-dimensional Coulomb branch, this amounts to a 
geometric $S_n$-quotient. In this note, we prove the conjecture on Coulomb 
branches with unitary nodes and, moreover, extend it to Coulomb branches with 
other classical groups. The results promote discrete $S_n$-quotients to a 
versatile tool in the study of Coulomb branches.
\end{abstract}

\end{center}

\end{titlepage}
{\baselineskip=12pt
{\footnotesize
\tableofcontents
}
}
  \section{Introduction and motivation}
The study of Coulomb branches of $3$-dimensional $\Ncal=4$ gauge theories has 
been proven vital for the understanding of gauge theories with 
$8$ supercharges in $5$ and $6$ dimensions. A powerful tool for 
analysing 
Coulomb branches as algebraic varieties is the Hilbert series --- called 
monopole formula \cite{Cremonesi:2013lqa} --- of the associated chiral ring.
In the original $3$-dimensional set-up, the monopole formula has provided a 
large number of interesting results and geometric insights 
\cite{Cremonesi:2014xha,Cremonesi:2014uva,Cremonesi:2014kwa,Cremonesi:2014vla,
Hanany:2015hxa,
Hanany:2016ezz,Hanany:2016pfm, Hanany:2016gbz, 
Hanany:2017ooe,Cabrera:2016vvv,Cabrera:2017ucb,Cabrera:2017njm,Cheng:2017got,
Hanany:2017pdx, Hanany:2018uzt}, for instance in the geometry of nilpotent 
orbits.

The standard lore suggests that Higgs branches of theories with $8$ supercharges 
in dimensions $3$, $4$, $5$, and $6$ are classically exact. In $5$-dimensional 
$\Ncal=1$ theories, the Higgs branch 
$\MHiggs^{5d}\big|_{g=\infty}$ at infinite gauge coupling, however, grows as 
new massless degrees of freedom appear in the form of instanton operators. 
As the Higgs branch is still a hyper-Kähler space, 
$\MHiggs^{5d}\big|_{g=\infty}$ has a $3$-dimensional Coulomb branch 
counterpart, provided the global symmetry is large enough 
\cite{Cremonesi:2015lsa,Ferlito:2017xdq}. To be precise, this means that a 
$3$-dimensional $\Ncal=4$ gauge theory exists such that its Coulomb branch 
agrees with $\MHiggs^{5d}\big|_{g=\infty}$. Such a quiver is further realised 
in the study of 5-brane webs and 7-branes.

Similarly, $6$-dimensional $\Ncal=(1,0)$ theories exhibit a previously 
unappreciated rich phase structure of the Higgs branch as recent 
developments have shown \cite{Hanany:2018uhm,Hanany:2018vph}. 
As it turns out, many interesting effects on the $6$-dimensional Higgs 
branches can be described neatly by associated $3$-dimensional $\Ncal=4$ 
theories, 
whose Coulomb branches $\MCoulomb^{3d}$ agree with the $6$-dimensional 
Higgs branches $\MHiggs^{6d}$ as algebraic varieties. In particular, 
the $3$-dimensional quiver gauge theory is understood as a tool that captures 
the geometry of the moduli space.
Besides the small $E_8$-instanton transition \cite{Hanany:2018uhm}, another 
interesting 
phenomenon is \emph{discrete gauging} \cite{Hanany:2018vph}. 
For the latter, it is crucial to realise that the gauging of a discrete global 
symmetry $\Gamma$ on $\MHiggs^{6d}$ corresponds to a quotient of 
$\MCoulomb^{3d}$ by $\Gamma$. In other words, restriction to the 
$\Gamma$-invariant sector.

In this note we prove earlier conjectures and extend the concept of discrete 
quotients of $3$-dimensional Coulomb branches to other scenarios. 
To begin with, we recall two examples.
\paragraph{Symmetric products of ALE spaces.}
Consider $k$ D2 branes in the presence of $n$ D6 branes in flat space. The 
worldvolume theory on the D2 branes is a $3$-dimensional $\Ncal=4$ $\urm(k)$ 
gauge theory with one adjoint and $n$ fundamental hypermultiplets. The 
corresponding quiver theory is the A-type ADHM quiver
\begin{align}
T_{k,n}^{\text{A-type}}=
 \raisebox{-.5\height}{
 	\begin{tikzpicture}
	\tikzstyle{gauge} = [circle, draw];
	\tikzstyle{flavour} = [regular polygon,regular polygon sides=4, draw];
	\node (g1) [gauge,label=right:{$\urm(k)$}] {};
	\node (f1) [flavour,below of=g1,label=right:{$\surm(n)$}] {};
	\draw (g1)--(f1);
	\node (l1) [above right of=g1]{$\Adj$};
	\draw [-] (0.11,0.15) arc (-70:90:10pt);
	\draw [-] (-0.11,0.15) arc (250:90:10pt);
	\end{tikzpicture}
	}
\end{align}
such that the Higgs branch is the moduli space $\Mcal_{k,\surm(n),\C^2}$ of $k$ 
$\surm(n)$-instantons on $\C^2$. Moreover, 3d mirror 
symmetry predicts that the Coulomb branch is the symmetric product of $k$ copies 
of the $A_{n-1}$ singularity \cite{deBoer:1996mp,Porrati:1996xi}. In detail,
\begin{align}
 \MHiggs\left( T_{k,n}^{\text{A-type}}\right)
	= \Mcal_{k,\surm(n),\C^2}
	\qquad \text{and} \qquad
\MCoulomb \left( T_{k,n}^{\text{A-type}}\right)
	= \Sym{k}\left( \C^2 \slash \Z_n \right) \;.
\end{align}
We recall that a $3$-dimensional $\Ncal=4$ $\uo$ gauge theory with $n$ 
fundamentals has $\C^2 \slash \Z_n$ as Coulomb 
branch; hence, we may write
\begin{align}
 \MCoulomb \left(
 \raisebox{-.5\height}{
 	\begin{tikzpicture}
	\tikzstyle{gauge} = [circle, draw];
	\tikzstyle{flavour} = [regular polygon,regular polygon sides=4, draw];
	\node (g1) [gauge,label=right:{$\urm(k)$}] {};
	\node (f1) [flavour,below of=g1,label=right:{$\surm(n)$}] {};
	\draw (g1)--(f1);
	\node (l1) [above right of=g1]{$\Adj$};
	\draw [-] (0.11,0.15) arc (-70:90:10pt);
	\draw [-] (-0.11,0.15) arc (250:90:10pt);
	\end{tikzpicture}
	}
	\right)
	=
	\Sym{k} \left(\MCoulomb \left( 
	\raisebox{-.5\height}{
 	\begin{tikzpicture}
	\tikzstyle{gauge} = [circle, draw];
	\tikzstyle{flavour} = [regular polygon,regular polygon sides=4, draw];
	\node (g1) [gauge,label=right:{$\uo$}] {};
	\node (f1) [flavour,below of=g1,label=right:{$\surm(n)$}] {};
	\draw (g1)--(f1);
	\end{tikzpicture}
	}
	\right)\right) \; .
	\label{eq:Ex_U(k)}
\end{align}
These symmetry properties have been conjectured in two complementary studies: 
firstly, by computing the quantum corrections to the Coulomb branch metric in 
\cite{deBoer:1996mp} and, secondly, by computing the Coulomb branch Hilbert 
series in \cite{Cremonesi:2013lqa}.

Extending the above setting by an orientifold O6 plane changes the resulting 
$3$-dimensional $\Ncal=4$ worldvolume theory to an $\usprm(2k)$ gauge theory 
with one antisymmetric and $n$ fundamental hypermultiplets. The quiver theory is 
again an ADHM quiver 
\begin{align}
T_{k,n}^{\text{D-type}}=
	 \raisebox{-.5\height}{
 	\begin{tikzpicture}
	\tikzstyle{gauge} = [circle, draw];
	\tikzstyle{flavour} = [regular polygon,regular polygon sides=4, draw];
	\node (g1) [gauge,label=right:{$\usprm(2k)$}] {};
	\node (f1) [flavour,below of=g1,label=right:{$\sorm(2n)$}] {};
	\draw (g1)--(f1);
	\node (l1) [above right of=g1]{$\Lambda^{2}$};
	\draw [-] (0.11,0.15) arc (-70:90:10pt);
	\draw [-] (-0.11,0.15) arc (250:90:10pt);
	\end{tikzpicture}
	}
\end{align}
such that the Higgs branch is the moduli space of $k$ $\sorm(2n)$-instantons on 
$\C^2$. Again, 3d mirror symmetry predicts that the Coulomb branch 
is the symmetric product of $k$ copies of the $D_n$ singularity 
\cite{deBoer:1996mp,Porrati:1996xi}. In other words,
\begin{align}
 \MHiggs\left( T_{k,n}^{\text{D-type}} \right)
	= \Mcal_{k,\sorm(2n),\C^2}
	\qquad \text{and} \qquad
\MCoulomb \left(  T_{k,n}^{\text{D-type}} \right)
	= \Sym{k}\left( \C^2 \slash D_n \right) \; .
\end{align}
Recalling that the Coulomb branch of a $\surm(2)\cong \usprm(2)$ gauge theory  
with $n$ fundamentals is the $D_n$-singularity, one may conclude
\begin{align}
 \MCoulomb \left(
	 \raisebox{-.5\height}{
 	\begin{tikzpicture}
	\tikzstyle{gauge} = [circle, draw];
	\tikzstyle{flavour} = [regular polygon,regular polygon sides=4, draw];
	\node (g1) [gauge,label=right:{$\usprm(2k)$}] {};
	\node (f1) [flavour,below of=g1,label=right:{$\sorm(2n)$}] {};
	\draw (g1)--(f1);
	\node (l1) [above right of=g1]{$\Lambda^{2}$};
	\draw [-] (0.11,0.15) arc (-70:90:10pt);
	\draw [-] (-0.11,0.15) arc (250:90:10pt);
	\end{tikzpicture}
	}
	\right)
	=
	\Sym{k} \left(\MCoulomb \left( 
	 \raisebox{-.5\height}{
 	\begin{tikzpicture}
	\tikzstyle{gauge} = [circle, draw];
	\tikzstyle{flavour} = [regular polygon,regular polygon sides=4, draw];
	\node (g1) [gauge,label=right:{$\usprm(2)$}] {};
	\node (f1) [flavour,below of=g1,label=right:{$\sorm(2n)$}] {};
	\draw (g1)--(f1);
	\end{tikzpicture}
	}
	\right)\right) \; .
	\label{eq:Ex_Sp(k)}
\end{align}
Again, this has been conjectured in \cite{deBoer:1996mp} and 
\cite{Cremonesi:2013lqa} from different approaches.

Lastly, replacing the O6 plane by a hypothetical 
$\widetilde{\mathrm{O6}}^+$ plane \cite{Hanany:2000fq,Feng:2000eq} implies that 
the $3$-dimensional $\Ncal=4$ worldvolume theory turns into a $\sorm(2k+1)$ 
gauge theory with one symmetric and $n$ fundamental hypermultiplets. The quiver 
is given by
\begin{align}
T_{k,n}^{\text{D}'\text{-type}}=
	 \raisebox{-.5\height}{
 	\begin{tikzpicture}
	\tikzstyle{gauge} = [circle, draw];
	\tikzstyle{flavour} = [regular polygon,regular polygon sides=4, draw];
	\node (g1) [gauge,label=right:{$\sorm(2k+1)$}] {};
	\node (f1) [flavour,below of=g1,label=right:{$\usprm(2n)$}] {};
	\draw (g1)--(f1);
	\node (l1) [above right of=g1]{$\; \; \Sym{2}$};
	\draw [-] (0.11,0.15) arc (-70:90:10pt);
	\draw [-] (-0.11,0.15) arc (250:90:10pt);
	\end{tikzpicture}
	}
	\label{eq:Dn+3-type}
\end{align}
and it has been conjectured in \cite{Cremonesi:2013lqa} that the 
Coulomb branch is again the $k$-th symmetric product of a $D$-type singularity, 
i.e.
\begin{align}
 \MCoulomb\left(T_{k,n}^{\text{D}'\text{-type}} \right) = \Sym{k}\left( 
\C^2 \slash D_{n+3}\right)\; .
\end{align}
%
%
\paragraph{6d Higgs branches.}
Following \cite{Hanany:2018vph}, consider $n$ separated M5-branes on a $\C^2 
\slash \Z_k$ singularity. The $6$-dimensional $\Ncal=(1,0)$ worldvolume theory 
has $(n-1)$ tensor multiplets, a $\surm(k)^{n-1}$ gauge group and bifundamental 
matter determined by a linear quiver
\begin{align}
  Q_{n,k}^{\mathrm{A}}=\raisebox{-.5\height}{
 	\begin{tikzpicture}
	\tikzstyle{gauge} = [circle, draw];
	\tikzstyle{flavour} = [regular polygon,regular polygon sides=4, draw];
	\node (g1) [gauge,label=below:{$k$}] {};
	\node (g2) [gauge,right of=g1,label=below:{$k$}] {};
	\node (g3) [right of=g2] {$\ldots$};
	\node (g4) [gauge,right of=g3,label=below:{$k$}] {};
	\node (g5) [gauge,right of=g4,label=below:{$k$}] {};
	\node (f1) [flavour,above of=g1,label=above:{$k$}] {};
	\node (f5) [flavour,above of=g5,label=above:{$k$}] {};
	\draw (g1)--(g2) (g2)--(g3) (g3)--(g4) (g4)--(g5) (g1)--(f1) (g5)--(f5);
	\end{tikzpicture}
	}  
\end{align}
where all nodes are special unitary gauge or flavour nodes.
The corresponding $3$-dimensional $\Ncal=4$ quiver gauge theory for $n$ 
separated M5-branes is equipped with a \emph{bouquet} of $n$ nodes with $1$, 
i.e.
\begin{align}
  F_{n,k}^{\mathrm{A}}=\raisebox{-.5\height}{
 	\begin{tikzpicture}
	\tikzstyle{gauge} = [circle, draw];
	\tikzstyle{flavour} = [regular polygon,regular polygon sides=4, draw];
	\node (g1) [gauge,label=below:{$1$}] {};
	\node (g2) [gauge,right of=g1,label=below:{$2$}] {};
	\node (g3) [right of=g2] {$\ldots$};
	\node (g4) [gauge,right of=g3,label=below:{$k$}] {};
	\node (g5) [right of=g4] {$\ldots$};
	\node (g6) [gauge,right of=g5,label=below:{$2$}] {};
	\node (g7) [gauge,right of=g6,label=below:{$1$}] {};
	\node (g8) [gauge,above left of=g4,label=above left:{$1$}] {};
	\node (g9) [gauge,above right of=g4,label=above right:{$1$}] {};
	\node (g10) [above of=g4,label=above:{$n$}] {$\ldots$};
	\draw (g1)--(g2) (g2)--(g3) (g3)--(g4) (g4)--(g5) (g5)--(g6) (g6)--(g7)
	 (g4)--(g8) (g4)--(g9);
	\end{tikzpicture}
	}  
\end{align}
with all nodes being unitary gauge groups.
It is important to appreciate the global discrete $S_n$ symmetry present in the 
problem of $n$ identical objects. In particular, the Coulomb branch quiver has 
an apparent $S_n$ symmetry. Following \cite{Hanany:2018vph}, the system 
admits different phases, in which $n_i$ of the $n$ 
separated M5-branes become coincident at positions $x_i$. These 
phases are then obtained by gauging a discrete $\prod_i S_{n_i} \subset S_n$ 
global symmetry in the $6$-dimensional theory. Let us summarise the main 
conjectures of \cite{Hanany:2018vph}:
\begin{compactenum}[(i)]
 \item At infinite gauge coupling, the $3$-dimensional quiver gauge theory for 
$n$ coinciding M5-branes on a $A_{k-1}$-singularity is given by
\begin{align}
  I_{n,k}^{\mathrm{A}}=\raisebox{-.5\height}{
 	\begin{tikzpicture}
	\tikzstyle{gauge} = [circle, draw];
	\tikzstyle{flavour} = [regular polygon,regular polygon sides=4, draw];
	\node (g1) [gauge,label=below:{$1$}] {};
	\node (g2) [gauge,right of=g1,label=below:{$2$}] {};
	\node (g3) [right of=g2] {$\ldots$};
	\node (g4) [gauge,right of=g3,label=below:{$k$}] {};
	\node (g5) [right of=g4] {$\ldots$};
	\node (g6) [gauge,right of=g5,label=below:{$2$}] {};
	\node (g7) [gauge,right of=g6,label=below:{$1$}] {};
	\node (g8) [gauge,above of=g4,label=right:{$n$}] {};
	\node (l1) [above right of=g8]{$\Adj$};
	\draw [-] (3+0.11,1.15) arc (-70:90:10pt);
	\draw [-] (3-0.11,1.15) arc (250:90:10pt);
	\draw (g1)--(g2) (g2)--(g3) (g3)--(g4) (g4)--(g5) (g5)--(g6) (g6)--(g7)
	 (g4)--(g8) ;
	\end{tikzpicture}
	}  
\end{align}
where all nodes are unitary gauge groups.
\item The $6$-dimensional Higgs branches and $3$-dimensional Coulomb branches 
then satisfy the following relations:
\begin{align}
 \MHiggs^{6d}(Q_{n,k}^{\mathrm{A}})\big|_{g<\infty} &= 
\MCoulomb^{3d}(F_{n,k}^{\mathrm{A}}) \; , \qquad
 \MHiggs^{6d}(Q_{n,k}^{\mathrm{A}})\big|_{g=\infty} = 
\MCoulomb^{3d}(I_{n,k}^{\mathrm{A}}) \; , \\
\MCoulomb^{3d}(I_{n,k}^{\mathrm{A}}) &= \MCoulomb^{3d}(F_{n,k}^{\mathrm{A}}) 
\slash S_n \; .
\label{eq:Ex_6d_1}
\end{align}
\item Suppose a partition $\{n_i\}$ of $n$ describes that the $n$ M5-branes 
coincide in a pattern of $n_i$ coinciding branes at different locations. The 
case of all branes separated corresponds to $\{1^n\}$, while all of them 
coinciding corresponds to $\{n\}$, i.e.\ infinite gauge coupling. Then the 
associated $3$-dimensional quiver is conjectured to be
 \begin{align}
  F_{\{n_i\},k}^{\mathrm{A}}=\raisebox{-.5\height}{
 	\begin{tikzpicture}
	\tikzstyle{gauge} = [circle, draw];
	\tikzstyle{flavour} = [regular polygon,regular polygon sides=4, draw];
	\node (g1) [gauge,label=below:{$1$}] {};
	\node (g2) [gauge,right of=g1,label=below:{$2$}] {};
	\node (g3) [right of=g2] {$\ldots$};
	\node (g4) [gauge,right of=g3,label=below:{$k$}] {};
	\node (g5) [right of=g4] {$\ldots$};
	\node (g6) [gauge,right of=g5,label=below:{$2$}] {};
	\node (g7) [gauge,right of=g6,label=below:{$1$}] {};
	\node (g8) [gauge,above right of=g4,label=right:{$n_l$}] {};
	\node (g9) [gauge,above left of=g4,label=left:{$n_1$}] {};
	\node (l1) [above left of=g9]{$\Adj$};
	\draw [-] (2.3+0.11,0.85) arc (-70:90:10pt);
	\draw [-] (2.3+-0.11,0.85) arc (250:90:10pt);
	\node (l2) [above right of=g8]{$\Adj$};
	\draw [-] (3.7+0.11,0.85) arc (-70:90:10pt);
	\draw [-] (3.7-0.11,0.85) arc (250:90:10pt);
	\node (g10) [above of =g4] {$\ldots$};
	\draw (g1)--(g2) (g2)--(g3) (g3)--(g4) (g4)--(g5) (g5)--(g6) (g6)--(g7)
	 (g4)--(g8) (g4)--(g9);
	\end{tikzpicture}
	} 
\end{align}
with the relations
\begin{align}
 \MHiggs^{6d}(Q_{n,k}^{\mathrm{A}})\big|_{\{n_i\}} = 
\MCoulomb^{3d}(F_{\{n_i\},k}^{\mathrm{A}}) \; , 
\qquad
 \MCoulomb^{3d}(F_{\{n_i\},k}^{\mathrm{A}}) =  
\MCoulomb^{3d}(F_{\{1^n\},k}^{\mathrm{A}}) \slash \prod_i 
S_{n_i} \; .\label{eq:Ex_6d_2}
\end{align}
Here, $\prod_i S_{n_i}$ denotes the product of permutation groups which act on 
the sets of $n_i$ coincident M5-branes.
\end{compactenum}

Similarly, $n$ M5-branes on a $\C^2\slash D_k$ singularity have been considered 
in \cite{Hanany:2018uhm}. The $6$-dimensional $\Ncal=(1,0)$ worldvolume theory  
is comprised of $(2n-1)$ tensor multiplets as well as  gauge groups and 
hypermultiplets determined by the quiver
\begin{align}
Q_{n,k}^{\mathrm{D}}=
  \raisebox{-.5\height}{
 	\begin{tikzpicture}
	\tikzstyle{gauge} = [circle, draw];
	\tikzstyle{flavour} = [regular polygon,regular polygon sides=4, draw];
	\node (g1) [gauge,label={[rotate=-45]below right:$\usprm(2k{-}8)$}] {};
	\node (g2) [gauge,right 
of=g1,label={[rotate=-45]below right:$\orm(2k)$}] {};
	\node (g3) [gauge,right 
of=g2,label={[rotate=-45]below right:$\usprm(2k{-}8)$}] {};
	\node (g4) [right of=g3] {$\ldots$};
	\node (g5) [gauge,right 
of=g4,label={[rotate=-45]below right:$\usprm(2k{-}8)$}] {};
	\node (f1) [flavour,above of=g1,label=above:{$\orm(2k)$}] {};
	\node (f5) [flavour,above of=g5,label=above:{$\orm(2k)$}] {};
	\draw (g1)--(g2) (g2)--(g3) (g3)--(g4) (g4)--(g5) (g1)--(f1) (g5)--(f5);
	\end{tikzpicture}
	}  
\end{align}
such that there are $n$ $\usprm(2k-8)$ and $(n-1)$ $\orm(2k)$ gauge 
nodes in total. The associated $3$-dimensional quiver theory for $n$ 
coincident M5 branes, i.e.\ infinite gauge coupling in the $6$-dimensional 
theory, has been conjectured to be
\begin{align}
I_{n,k}^{\mathrm{D}}=
  \raisebox{-.5\height}{
 	\begin{tikzpicture}
	\tikzstyle{gauge} = [circle, draw];
	\tikzstyle{flavour} = [regular polygon,regular polygon sides=4, draw];
\node (g1) [gauge,label={[rotate=-45]below right:$\orm(2)$}] {};
\node (g2) [gauge,right of=g1,label={[rotate=-45]below right:$\usprm(2)$}] {};
\node (g3) [gauge,right of=g2,label={[rotate=-45]below right:$\orm(4)$}] {};
\node (g4) [gauge,right of=g3,label={[rotate=-45]below right:$\usprm(4)$}] {};
\node (g5) [right of=g4] {$\ldots$};
\node (g6) [gauge,right of=g5,label={[rotate=-45]below right:$\orm(2k{-}2)$}] 
{};
\node (g7) [gauge,right of=g6,label={[rotate=-45]below right:$\usprm(2k{-}2)$}] 
{};
\node (g8) [gauge,right of=g7,label={[rotate=-45]below right:$\orm(2k)$}] {};
\node (g9) [gauge,right of=g8,label={[rotate=-45]below right:$\usprm(2k{-}2)$}] 
{};
\node (g10) [gauge,right of=g9,label={[rotate=-45]below right:$\orm(2k{-}2)$}] 
{};
\node (g11) [right of=g10] {$\ldots$};
\node (g12) [gauge,right of=g11,label={[rotate=-45]below 
right:$\usprm(4)$}] {};
\node (g13) [gauge,right of=g12,label={[rotate=-45]below right:$\orm(4)$}] 
{};
\node (g14) [gauge,right of=g13,label={[rotate=-45]below 
right:$\usprm(2)$}] {};
\node (g15) [gauge,right of=g14,label={[rotate=-45]below right:$\orm(2)$}] 
{};
\node (g16) [gauge,above of=g8,label=right:{$\usprm(2n)$}] {};
	\node (l1) [above right of=g16]{$\Lambda^2$};
	\draw [-] (7+0.11,1.15) arc (-70:90:10pt);
	\draw [-] (7-0.11,1.15) arc (250:90:10pt);
	\draw (g1)--(g2) (g2)--(g3) (g3)--(g4) (g4)--(g5) (g5)--(g6) (g6)--(g7) 
(g7)--(g8) (g8)--(g9) (g9)--(g10) (g10)--(g11) (g11)--(g12) (g12)--(g13) 
(g13)--(g14) (g14)--(g15) (g8)--(g16);
	\end{tikzpicture}
	}  
\end{align}
where $\Lambda^2$ denotes the traceless rank-2 antisymmetric representation of 
$\usprm(2n)$. The theories are related via
\begin{align}
 \MHiggs^{6d}(Q_{n,k}^{\mathrm{D}})\big|_{g=\infty} = 
\MCoulomb^{3d}(I_{n,k}^{\mathrm{D}}) \; .
\end{align}
One can argue, as shown below, that 
$\MCoulomb^{3d}(I_{n,k}^{\mathrm{D}})$ is the $S_n$-quotient of the Coulomb 
branch of 
\begin{align}
F_{n,k}^{\mathrm{D}}=
  \raisebox{-.5\height}{
 	\begin{tikzpicture}
	\tikzstyle{gauge} = [circle, draw];
	\tikzstyle{flavour} = [regular polygon,regular polygon sides=4, draw];
\node (g1) [gauge,label={[rotate=-45]below right:$\orm(2)$}] {};
\node (g2) [gauge,right of=g1,label={[rotate=-45]below right:$\usprm(2)$}] {};
\node (g3) [gauge,right of=g2,label={[rotate=-45]below right:$\orm(4)$}] {};
\node (g4) [gauge,right of=g3,label={[rotate=-45]below right:$\usprm(4)$}] {};
\node (g5) [right of=g4] {$\ldots$};
\node (g6) [gauge,right of=g5,label={[rotate=-45]below right:$\orm(2k{-}2)$}] 
{};
\node (g7) [gauge,right of=g6,label={[rotate=-45]below right:$\usprm(2k{-}2)$}] 
{};
\node (g8) [gauge,right of=g7,label={[rotate=-45]below right:$\orm(2k)$}] {};
\node (g9) [gauge,right of=g8,label={[rotate=-45]below right:$\usprm(2k{-}2)$}] 
{};
\node (g10) [gauge,right of=g9,label={[rotate=-45]below right:$\orm(2k{-}2)$}] 
{};
\node (g11) [right of=g10] {$\ldots$};
\node (g12) [gauge,right of=g11,label={[rotate=-45]below 
right:$\usprm(4)$}] {};
\node (g13) [gauge,right of=g12,label={[rotate=-45]below right:$\orm(4)$}] 
{};
\node (g14) [gauge,right of=g13,label={[rotate=-45]below 
right:$\usprm(2)$}] {};
\node (g15) [gauge,right of=g14,label={[rotate=-45]below right:$\orm(2)$}] 
{};
\node (g16) [gauge,above left of=g8,label=left:{$\usprm(2)$}] {};
\node (g17) [above of=g8,label=above:{$n$}] {$\ldots$};
\node (g18) [gauge,above right of=g8,label=right:{$\usprm(2)$}] {};
	\draw (g1)--(g2) (g2)--(g3) (g3)--(g4) (g4)--(g5) (g5)--(g6) (g6)--(g7) 
(g7)--(g8) (g8)--(g9) (g9)--(g10) (g10)--(g11) (g11)--(g12) (g12)--(g13) 
(g13)--(g14) (g14)--(g15) (g8)--(g16) (g8)--(g18);
	\end{tikzpicture}
	}  
\end{align}
which is a quiver with a \emph{bouquet} of $n$ nodes of $\usprm(2)$. 
Physically, $F_{n,k}^{\mathrm{D}}$ captures the phase in which all 
$n$ M5-branes are separated. The discrete gauging on the Higgs branch is 
reflected by the relation of the Coulomb branches
\begin{align}
 \MCoulomb^{3d}(I_{n,k}^{\mathrm{D}}) &= \MCoulomb^{3d}(F_{n,k}^{\mathrm{D}}) 
\slash S_n \; .
\end{align}
As shown below, one can generalise the setting to the analogue of 
\eqref{eq:Ex_6d_2}. 

\paragraph{Outline.}
From the above examples, there is an apparent action of an $S_n$ group on a 
Coulomb branch quiver, which appears to be a local operation on the quiver. 
These examples serve as guideline to prove exact statements on the Coulomb 
branch Hilbert series upon the action of an $S_n$ group. 

The remainder is organised as follows: the generalisations of the examples 
discussed in the introduction are the focus of Section \ref{sec:A_and_D-type}. 
In detail, the generalisation to an arbitrary quiver coupled to either a 
bouquet of $\uo$, $\usprm(2)$, or $\sorm(3)$ nodes is considered and the 
statements of discrete $S_n$-quotients are proven on the level of the monopole 
formula. In Section \ref{sec:applications}, 
applications to other types of bouquets, composed of (different) $\usprm(2)$, 
$\sorm(3)$, and $\orm(2)$ nodes, are considered and proven. Thereafter, 
Section \ref{sec:conclusion} 
summarises and concludes. Appendix \ref{app:all} provides some 
background on the cycle index and a proof of an auxiliary identity.

As a remark, a  complementary perspective of discrete gauging and its 
manifestation as discrete quotients on Coulomb branches is presented in 
\cite{Hanany:2018dvd}.
  \section{A and D-type}
\label{sec:A_and_D-type}
This section focuses on generic good $3$-dimensional $\Ncal=4$ quiver 
gauge theories 
that are coupled to bouquets of either $\uo$, $\usprm(2)$, or $\sorm(3)$ nodes. 
Upon $S_n$-quotient, the bouquet is replaced by a single $\urm(n)$, 
$\usprm(2n)$, or $\sorm(2n+1)$ node supplemented by an additional 
hypermultiplet that transforms as in the corresponding ADHM quiver. 
\subsection{A-type --- U(1)-bouquet}
\label{sec:A-type}
Taking \eqref{eq:Ex_U(k)} as well as \eqref{eq:Ex_6d_1} and \eqref{eq:Ex_6d_2} 
as motivation, one can generalise the statement to a generic quiver with one 
(or many) bouquet(s) attached and provide a proof on the level of the monopole 
formula.

Consider an arbitrary quiver, denoted by $\bullet$, coupled to either a 
$\urm(n)$ gauge node with one additional adjoint hypermultiplet or a bouquet 
of $n$ $\uo$ nodes. I.e. define the two quiver theories
\begin{align}
 T_{\{n\},\bullet} = 
 \raisebox{-.5\height}{
 \begin{tikzpicture}
	\tikzstyle{gauge} = [circle, draw];
	\tikzstyle{flavour} = [regular polygon,regular polygon sides=4, draw];
	\tikzstyle{extra} = [circle, fill=black, draw];
	\node (e1) [extra] {};
	\node (g1) [gauge,above of=e1,label=right:{$\urm(n)$}] {};;
	\node (g2) [above right of=g1]{$\Adj$};
	\draw [-] (0.11,1.15) arc (-70:90:10pt);
	\draw [-] (-0.11,1.15) arc (250:90:10pt);
	\draw 	(g1)--(e1);
	\end{tikzpicture}
	}
	\qquad \text{and}  \qquad
 T_{\{1^n\},\bullet}=
 \raisebox{-.5\height}{
 \begin{tikzpicture}
	\tikzstyle{gauge} = [circle, draw];
	\tikzstyle{flavour} = [regular polygon,regular polygon sides=4, draw];
	\tikzstyle{extra} = [circle, fill=black, draw];
	\node (e1) [extra] {};
	\node (g1) [gauge,above left of =e1,label=above left:{$\uo$}] {};
	\node (g2) [gauge,above right of =e1,label=above right:{$\uo$}] {};
	\node (g3) [above of = e1,label=above:{$n$}] {$\cdots$};
		\draw 	(g1)--(e1)  (g2)--(e1);
	\end{tikzpicture}
	} \;.
	\label{eq:A-type_quiver}
\end{align}
To be precise, the $\urm(n)$ as well as all of the $\uo$ nodes couple to the 
same \emph{single} node in $\bullet$ via bifundamental matter. Viewed 
from the $\urm(n)$ or $\uo$ nodes, the quiver $\bullet$ is considered as 
providing background charges $\vec{k}=(k_1,\ldots,k_s)$ for some $s\in \NN$, 
i.e.\ the magnetic charges from the single node they couple to. 
To see this, consider this single node in $\bullet$ as flavour node with 
background charges $\vec{k}$ as, for example, in 
\cite{Cremonesi:2014kwa,Cremonesi:2014vla,Hanany:2018uzt}. Thus, there are two 
Hilbert series to compute: (i) the monopole formula $H(t,\vec{k})$ of $\bullet$ 
with the single node turned into a flavour node with fluxes $\vec{k}$, and (ii) 
the monopole formula of $T_{\{n\},\Box}$ (or $T_{\{1^n\},\Box}$) where the 
flavour node $\Box$ provides fluxes $\vec{k}$. The Hilbert series of 
$T_{\{n\},\bullet}$ (or $T_{\{1^n\},\bullet}$) can be obtained via gluing the 
Hilbert series $H(t,\vec{k})$ with that of $T_{\{n\},\Box}$ (or 
$T_{\{1^n\},\Box}$) along the common flavour node, which turns it into a gauge 
node. Since the Hilbert series with background charges for $\bullet$ is same in 
both cases and is not affected by the $S_n$-quotient, it will henceforth be 
ignored.
Now, the above conjecture \eqref{eq:Ex_6d_1} is generalised by 
\begin{myProp}
Let the quiver gauge theories $T_{\{n\},\bullet}$ and $T_{\{1^n\},\bullet}$ be 
as defined in \eqref{eq:A-type_quiver}, then their Coulomb branches satisfy
\begin{align}
\MCoulomb \left(
T_{\{n\},\bullet}
	\right)
	=
	\MCoulomb \left( T_{\{1^n\},\bullet} \right)
	 \slash S_n \; .
	\label{eq:claim_A-type} 
\end{align} 
\label{prop:A-type}
\end{myProp}
\paragraph{Preliminaries.}
In order to prove Proposition \ref{prop:A-type}, one defines
\begin{align}
 f(t,z)= \HS_{\MCoulomb} \left(
 \raisebox{-.5\height}{
  \begin{tikzpicture}
	\tikzstyle{gauge} = [circle, draw];
	\tikzstyle{flavour} = [regular polygon,regular polygon sides=4, draw];
	\tikzstyle{extra} = [circle, fill=black, draw];
	\node (e1) [extra] {};
	\node (g1) [gauge,above of =e1,label=above:{$\uo$}] {};
	\draw (g1)--(e1)  ;
	\end{tikzpicture} 
	}
 \right)  \equiv 
 \HS_{\MCoulomb \left(T_{\{1\},\bullet} \right)} 
\end{align}
as Hilbert series of the $3$-dimensional $\Ncal=4$ $\uo$ gauge theory with 
background charges $\vec{k}$. In detail, the 
conformal dimension and dressing factor read
\begin{align}
 \Delta(q;\vec{k}) = \frac{1}{2}  |q-\vec{k}| \; , \qquad
 P(t;q) = \frac{1}{1-t} 
\end{align}
for $q\in \Z$.
Then \eqref{eq:claim_A-type} can be expressed via the following generating 
series:
\begin{align}
 F[\nu;t,z]&=\PE [\nu \cdot f(t,z)]  
 = \sum_{n=0}^{\infty} \nu^n\, \HS_n(t,z)
\end{align}
such that Proposition \ref{prop:A-type} becomes
\begin{align} 
\HS_n(t,z) \equiv \HS_{\MCoulomb \left(T_{\{n\},\bullet} \right)} 
  \stackrel{\text{Prop.\ \ref{prop:A-type}}}{=} 
  \left. \frac{1}{n!} \frac{\diff^n }{\diff \nu^n} \PE \left[\nu \cdot
  \HS_{\MCoulomb \left(T_{\{1\},\bullet} \right)}  \right]  
\right|_{\nu=0} 
\equiv \HS_{\MCoulomb \left(T_{\{1^n\},\bullet} \right) \slash S_n } 
 \; .
 \label{eq:HS_explicit_A}
\end{align}
In order to compute $\HS_n(t,z) $ explicitly from the symmetrisation of 
$f(t,z)$, one employs the cycle index \eqref{eq:cycle_index_def}. 
 
To compare the result, recall the ingredients for the monopole formula 
of an $\urm(n)$ gauge node with one adjoint hypermultiplet and background 
charges. The conformal dimension reads
\begin{align}
 \Delta(q_1,\ldots,q_n;\vec{k}) = \frac{1}{2} \sum_{i=1}^n |q_i-\vec{k}| 
 = \sum_{i=1}^n \Delta(q_i;\vec{k})
\end{align}
wherein the contributions from the adjoint hypermultiplet cancel the vector 
multiplet contributions entirely. The magnetic charges appearing in the 
monopole formula are ordered $q_1\geq q_2\geq \ldots \geq q_n$. The $\urm(n)$ 
dressing factors have been 
defined in \cite{Cremonesi:2013lqa}. The shorthand notation $|q_i-\vec{k}| 
\equiv \sum_{l=1}^s |q_i -k_l|$ summarises the contributions from the magnetic 
charges $k_l$ of the single node in $\bullet$ the $\urm(n)$ couples to via 
bifundamental matter.
\paragraph{Proof.}
The recursive formula \eqref{eq:cycle_index_recursive} suggests to prove 
\eqref{eq:claim_A-type} by induction. One explicitly verifies the claim for 
$n=1,2$; thereafter one proceeds to $\HS_n(t,z)$ with general $n$, i.e.\
\begin{align}
 \HS_n(t,z) = \frac{1}{n} \sum_{k=1}^n a_k \cdot \HS_{n-k}(t,z)  \, , \quad a_k 
= f(t^k,z^k) \; ,
\label{eq:HS_recursive}
\end{align}
assuming the validity for all $\HS_{k}(t,z)$ with $k<n$.
To begin with, one verifies the base case:
\begin{compactenum}[(i)]
 \item $n=1$: trivial. Returns the $\uo$ case.
\item $n=2$: The two contributions read
\begin{align}
 a_1^2 &= \frac{1}{(1-t)^2} \sum_{q_1,q_2} z^{q_1+q_2} t^{\Delta(q_1) + 
\Delta(q_2)} \notag \\
 &= \frac{2}{(1-t)^2} \sum_{q_1>q_2} z^{q_1+q_2} t^{\Delta(q_1) + \Delta(q_2)} 
 +\frac{1}{(1-t)^2} \sum_{q_1=q_2} z^{q_1+q_2} t^{\Delta(q_1) + \Delta(q_2)} 
\;, \\
a_2 &= \frac{1}{1-t^2} \sum_{q} z^{2q} t^{2\Delta(q)} \; .
\end{align}
Combining both, one obtains
\begin{align}
 \HS_2(t,z) &= \frac{1}{(1-t)^2} \sum_{q_1>q_2} z^{q_1+q_2} t^{\Delta(q_1)  + 
\Delta(q_2)} 
 +\frac{1}{2} \left( \frac{1}{(1-t)^2} +
  \frac{1}{1-t^2} \right)\sum_{q_1=q_2} z^{q_1+q_2} t^{\Delta(q_1) + 
\Delta(q_2)} \notag \\
&= \frac{1}{(1-t)^2} \sum_{q_1>q_2} z^{q_1+q_2} t^{\Delta(q_1) + \Delta(q_2)} 
 +  \frac{1}{(1-t)(1-t^2)} \sum_{q_1=q_2} z^{q_1+q_2} t^{\Delta(q_1) + 
\Delta(q_2)}  \; ,
\end{align}
which coincides with the monopole formula for the quiver $\bullet$ coupled to a 
$\urm(2)$ gauge node with an adjoint hypermultiplet.
\end{compactenum}
%
%
Thereafter, one proceeds with the inductive step $(n-1) \to n$ 
for \eqref{eq:HS_recursive}. The strategy of the proof is to consider the 
different contributions for the distinct summation regions of the magnetic 
charges $q_i$ in detail and show that these agree with the monopole formula of 
$T_{\{n\},\bullet}$.
\begin{compactenum}[(i)]
 \item $q_1 > q_2 > \ldots > q_n$ can only originate from one term:   $a_1 
\HS_{n-1}$, in which one denotes the magnetic charge in $a_1$ by $q$ and 
those of $\HS_{n-1}$ by $q_i$, $i=1,\ldots,n-1$. Then there are exactly $n$ 
contributing cases: 
 \begin{align}
 \begin{aligned}
q > q_1 > \ldots > q_{n-1} \;,\\
q_1 > q > q_2 > \ldots \;,\\
\ldots \;,\\
q_1 > \ldots > q > q_{n-1} \;,\\
q_1 > \ldots > q_{n-1} > q \;,
\end{aligned}
 \end{align}
 but these can all be relabelled to a single case. Then one finds
 \begin{align}
  \frac{1}{n} a_1 \HS_{n-1} \supset \frac{1}{(1-t)^n} \sum_{q_1> \ldots > q_n} 
  z^{\sum_{j=1}^n q_j}
t^{\sum_{j=1}^n \Delta(q_j)} \
 \end{align}
 and observes the dressing factor for a residual $\uo^n$ gauge symmetry, which 
is in fact the correct stabiliser of $q_1 > q_2 > \ldots > q_n$ in $\urm(n)$.
\item $q_1 > q_2 > \ldots  q_{i}> q_{i+1}=\ldots = q_{i+l} > \ldots >q_n $. The 
relevant contributions can only arise from $a_1 \HS_{n-1}$ to $a_l 
\HS_{n-l}$. Then $a_1 \HS_{n-1}$ has two different contributions: firstly,
\begin{align}
a_1 \HS_{n-1} \supset
  \sum_q \frac{1}{1-t}  z^q  t^{ \Delta(q)}
  \; \sum_{ \substack{l\text{ equal } 
q_i \\ \text{out of }n-1} } \frac{1}{\prod_{a=1}^l (1-t^a) } 
\frac{1}{(1-t)^{n-1-l}}  z^{\sum_{j=1}^{n-1} q_j} 
t^{\sum_{j=1}^{n-1} \Delta(q_j)},
\end{align}
but recall that the $q_i$ in $\HS_{n-1}$ are already ordered. Then there are 
exactly $(n-l)$ possible ways to arrange $q$ in between the $q_i$. 
However, a simple relabelling makes them all identical and one obtains:
\begin{align}
a_1 \HS_{n-1} \supset
 (n-l) \cdot  \sum_{ \substack{l\text{ equal } 
q_i \\ \text{out of }n} } \frac{1}{\prod_{a=1}^l (1-t^a) } 
\frac{1}{(1-t)^{n-l}} \cdot z^{\sum_{j=1}^n q_j}  t^{\sum_{j=1}^n 
\Delta(q_j)} \; .
\end{align}
Secondly, there is the contribution where $(l-1)$ $q_i$ are equal in 
$\HS_{n-1}$ 
and one has to align the $q$ from $a_1$ with those $(l-1)$ equal magnetic 
charges. This yields precisely one case
\begin{align}
a_1 \HS_{n-1} \supset
   \sum_{ \substack{l\text{ equal } 
q_i \\ \text{out of }n} } \frac{1}{\prod_{a=1}^{l-1} (1-t^a) } 
\frac{1}{(1-t)^{n-l}} \frac{1}{(1-t)} \cdot
z^{\sum_{j=1}^n q_j} t^{\sum_{j=1}^n \Delta(q_j)} \; .
\end{align}
Similarly, there exists exactly one matching contribution for $a_j \HS_{n-j}$, 
where the $q$ from $a_j$ has to match the $(l-j)$ equal $q_i$ from $\HS_{n-j}$. 
One obtains
\begin{align}
a_j \HS_{n-j} \supset
    \sum_{ \substack{l\text{ equal } 
q_i \\ \text{out of }n} } \frac{1}{\prod_{a=1}^{l-j} (1-t^a) } 
\frac{1}{(1-t)^{n-l}} \frac{1}{(1-t^j)} \cdot 
z^{\sum_{j=1}^n q_j}
t^{\sum_{j=1}^n \Delta(q_j)} \; .
\end{align}
The total contribution becomes
\begin{align}
 \sum_{j=1}^l a_j \HS_{n_j}(t) &\supset
  \left(
 \frac{(n-l)}{\prod_{a=1}^l (1-t^a) } 
\frac{1}{(1-t)^{n-l}}
 +\sum_{j=1}^l
 \frac{1}{\prod_{a=1}^{l-j} (1-t^a) } 
\frac{1}{(1-t)^{n-l}} \frac{1}{(1-t^j)} 
\right) \notag \\
&\qquad \qquad \qquad \qquad \qquad\qquad \cdot
 \sum_{ \substack{l\text{ equal } 
q_i \\ \text{out of }n} } 
z^{\sum_{j=1}^n q_j}  t^{\sum_{j=1}^n \Delta(q_j)}  
\notag\\
&=\frac{1}{\prod_{a=1}^l (1-t^a) } 
\frac{1}{(1-t)^{n-l}} \left(
 n -l + 
 Q_l(t)   
\right)
 \sum_{ \substack{l\text{ equal } 
q_i \\ \text{out of }n} }   
z^{\sum_{j=1}^n q_j} t^{\sum_{j=1}^n \Delta(q_j)}  
\\
&\qquad \text{with} \qquad 
Q_l(t)\coloneqq  \sum_{j=1}^l 
 \frac{1}{(1-t^j)} 
 \prod_{a=l-j+1}^{l} (1-t^a) \; .
\end{align}
As proven in Appendix \ref{app:q-theory}, $Q_l(t)$ satisfies
\begin{align}
 Q_l(t) =l \qquad \forall t \; .
\end{align}
Such that one obtains the contribution:
\begin{align}
 \frac{1}{n} \sum_{j=1}^l a_j \HS_{n_j}(t) \supset \frac{1}{\prod_{a=1}^l 
(1-t^a) } 
\frac{1}{(1-t)^{n-l}} 
 \sum_{ \substack{l\text{ equal } 
q_i \\ \text{out of }n} } 
z^{\sum_{j=1}^n q_j} t^{\sum_{j=1}^n \Delta(q_j)}
\end{align}
and one recognises the correct dressing factor for the residual $S(\urm(l) 
\times \uo^{n-l}) $ gauge symmetry of $l$ equal magnetic charges.
\item Now, one can easily generalise to any partition $(l_1,\ldots,l_p)$, 
$\sum_{j=1}^i l_i=n$ (not necessarily ordered) that describes the set-up of 
\begin{align}
 q_1=\ldots=q_{l_1} > 
 q_{l_1 +1} = \ldots =q_{l_1+l_2}>
 \ldots >
 q_{l_1+l_2+\ldots+l_{p-1}+1}=\ldots =q_{l_1+l_2+\ldots+l_p} \,.
\end{align}
The total contribution becomes
\begin{align}
 \frac{1}{n} \sum_{j=1}^{\max{(l_i)}} a_j \HS_{n_j}(t) &\supset 
 \frac{1}{n} \frac{1}{\prod_{j=1}^p \prod_{a_j=1}^{l_j} (1-t^a_j) } 
 \left(
 n + \sum_{j=1}^p  \underbrace{ \left(  
 Q_{l_j}(t)    -l_j \right) }_{=0} 
\right)
 \sum_{ \substack{l\text{ equal } 
q_i \\ \text{out of }n} } 
z^{\sum_{j=1}^n q_j} t^{\sum_{j=1}^n \Delta(q_j)}  \notag \\
 &=
  \frac{1}{\prod_{j=1}^p \prod_{a_j=1}^{l_j} (1-t^a_j) }
 \sum_{ \substack{l\text{ equal } 
q_i \\ \text{out of }n} } 
z^{\sum_{j=1}^n q_j} t^{\sum_{j=1}^n \Delta(q_j)} \;,
\end{align}
which is the correct contribution with a dressing factor reflecting the 
residual $S(\prod_{j=1}^p \urm(l_j))$ gauge symmetry. Again, the factor $n$ is 
cancelled by the $\frac{1}{n}$ pre-factor in the cycle index.
\end{compactenum}
Consequently, one has addressed all possible $\{q_i\}$, $i=1,\ldots,n$, 
summation regions that appear in \eqref{eq:HS_recursive} and, most importantly, 
one has proven that these correspond exactly to the definition of the fully 
refined monopole formula for a $\urm(n)$ gauge node with one adjoint 
hypermultiplet and background charges.  This concludes the proof of Proposition 
\ref{prop:A-type}.
\paragraph{Comments.}
The proof shows that given the Coulomb branch of an arbitrary quiver $\bullet$ 
with a $\uo$-bouquet of size $n$, one may quotient by $S_n$. The result is the 
same 
as the Coulomb branch of $\bullet$ coupled to a $\urm(n)$-node with one 
additional adjoint hypermultiplet.
From the nature of the proof, i.e.\ the $S_n$-quotient is a local operation on 
the Coulomb branch, there exist two immediate corollaries:
\begin{compactenum}[(i)]
 \item Similarly to \eqref{eq:Ex_6d_2}, one can consider a generic partition 
$\{n_i\}$ of $n$ which corresponds to the quotient by $\prod_i S_{n_i}$ on 
$T_{\{1^n\},\bullet}$. Since $\bullet$ has been arbitrary, one can repeat the 
proof by subdividing the size $n$ bouquet, focusing on the sub-bouquet of size 
$n_i$, while treating the remaining $\uo$-nodes as part of the background 
quiver. In other word, the quiver
\begin{align}
 T_{\{n_i\},\bullet}=
 \raisebox{-.5\height}{
 \begin{tikzpicture}
	\tikzstyle{gauge} = [circle, draw];
	\tikzstyle{flavour} = [regular polygon,regular polygon sides=4, draw];
	\tikzstyle{extra} = [circle, fill=black, draw];
	\node (e1) [extra] {};
	\node (g1) [gauge,above left of =e1,label= left:{$\urm(n_1)$}] {};
	\node (g2) [gauge,above right of =e1,label= right:{$\urm(n_l)$}] {};
	\node (g3) [above of = e1] {$\cdots$};
	\draw 	(g1)--(e1)  (g2)--(e1);
	\node (l1) [above left of=g1]{$\Adj$};
	\draw [-] (-0.7+0.11,0.85) arc (-70:90:10pt);
	\draw [-] (-0.7+-0.11,0.85) arc (250:90:10pt);
	\node (l2) [above right of=g2]{$\Adj$};
	\draw [-] (0.7+0.11,0.85) arc (-70:90:10pt);
	\draw [-] (0.7-0.11,0.85) arc (250:90:10pt);
	\end{tikzpicture}
	}
\end{align}
has a Coulomb branch which satisfies
\begin{align}
 \MCoulomb\left(T_{\{n_i\},\bullet} \right) = 
\MCoulomb\left(T_{\{1^n\},\bullet} \right) \slash\prod_i S_{n_i} \; .
\end{align}
 \item Furthermore, one may consider quivers where multiple bouquets are 
attached to different nodes. Then one can repeat the process of discrete 
quotients to any of the bouquets individually, as the operation is entirely 
local. 
\end{compactenum}

%
%
\subsection{D-type --- USp(2)-bouquet}
\label{sec:D-type}
Next, one can generalise \eqref{eq:Ex_Sp(k)} by considering an arbitrary quiver 
coupled to one (or many) bouquet(s) of $\usprm(2)\cong \surm(2)$ gauge nodes 
and provide a proof at the level of the monopole formula.

Again, consider an arbitrary quiver, labelled by $\bullet$, coupled to either 
an $\usprm(2n)$ gauge node with one additional anti-symmetric hypermultiplet or 
a $\usprm(2)$-bouquet of size $n$. Again, the following notation is employed:
\begin{align}
 T_{\{n\},\bullet} = 
 \raisebox{-.5\height}{
 \begin{tikzpicture}
	\tikzstyle{gauge} = [circle, draw];
	\tikzstyle{flavour} = [regular polygon,regular polygon sides=4, draw];
	\tikzstyle{extra} = [circle, fill=black, draw];
	\node (e1) [extra] {};
	\node (g1) [gauge,above of=e1,label=right:{$\usprm(2n)$}] {};
	\node (g2) [above right of=g1]{$\Lambda^2$};
	\draw [-] (0.11,1.15) arc (-70:90:10pt);
	\draw [-] (-0.11,1.15) arc (250:90:10pt);
	\draw 	(g1)--(e1);
	\end{tikzpicture}
	}
	\qquad \text{and}  \qquad
 T_{\{1^n\},\bullet}=
 \raisebox{-.5\height}{
 \begin{tikzpicture}
	\tikzstyle{gauge} = [circle, draw];
	\tikzstyle{flavour} = [regular polygon,regular polygon sides=4, draw];
	\tikzstyle{extra} = [circle, fill=black, draw];
	\node (e1) [extra] {};
	\node (g1) [gauge,above left of =e1,label=above left:{$\usprm(2)$}] {};
	\node (g2) [gauge,above right of =e1,label=above right:{$\usprm(2)$}] 
{};
	\node (g3) [above of = e1,label=above:{$n$}] {$\cdots$};
		\draw 	(g1)--(e1)  (g2)--(e1);
	\end{tikzpicture}
	} \;.
	\label{eq:D-type_quiver}
\end{align}
As above, the $\usprm(2n)$ as well as all of the $\usprm(2)$ nodes couple to 
the same \emph{single} node in $\bullet$ via bifundamental matter. From the 
$\usprm(2n)$ or $\usprm(2)$ point of view, the quiver $\bullet$ 
contributes background charges $\vec{k}=(k_1,\ldots,k_s)$ for some 
$s\in \NN$, i.e.\ the magnetic charges from the single node they couple to. All 
other contributions from $\bullet$ could be summarised in a function 
of the fugacity, which is not affected by the $S_n$-quotient and is henceforth 
ignored, cf. the discussion below \eqref{eq:A-type_quiver}.
\begin{myProp}
 Let the theories $ T_{\{n\},\bullet}$ and $T_{\{1^n\},\bullet}$ be as defined 
in \eqref{eq:D-type_quiver}, then the Coulomb branches satisfy
\begin{align}
\MCoulomb \left( T_{\{n\},\bullet}\right)
	=
\MCoulomb \left( T_{\{1^n\},\bullet} \right) \slash S_n \; .
\end{align}
\label{prop:D-type}
\end{myProp}
\paragraph{Preliminaries.}
To begin with, define the basic ingredient:
\begin{align}
 f(t)= \HS_{\MCoulomb} \left(
 \raisebox{-.5\height}{
  \begin{tikzpicture}
	\tikzstyle{gauge} = [circle, draw];
	\tikzstyle{flavour} = [regular polygon,regular polygon sides=4, draw];
	\tikzstyle{extra} = [circle, fill=black, draw];
	\node (e1) [extra] {};
	\node (g1) [gauge,above of =e1,label=above:{$\usprm(2)$}] {};
	\draw 	(g1)--(e1)  ;
	\end{tikzpicture} 
	}
 \right)
 \equiv \HS_{\MCoulomb(T_{\{1\},\bullet})}
 \,,
\end{align}
which is the Coulomb branch Hilbert series in the presence of background 
charges. Note that there is no extra topological fugacity for $\usprm(2)$.
The relevant conformal dimension is
\begin{align}
 \Delta(q;\vec{k})=\frac{1}{2}(|q-\vec{k}|+|q+\vec{k}|) - 2 |q|
\end{align}
for the magnetic charge $q \in \NN$ and background fluxes $\vec{k}$.
The dressing factors associated to $\usprm(2)$ are 
\begin{align}
 P(t,q) = 
 \begin{cases}
  \frac{1}{1-t} & ,\; q>0 \; ,\\ 
  \frac{1}{1-t^2} & ,\; q=0 \; .
 \end{cases}
\end{align}
The $\usprm(2n)$ gauge node with a hypermultiplet transforming in 
$\Lambda^2([1,0,\ldots,0])$ has the following conformal dimension:
\begin{align}
 \Delta(q_1,\ldots, q_n;\vec{k}) = 
 \frac{1}{2} \sum_{i=1}^n \left( |q_i -\vec{k}| + |q_i +\vec{k}| \right) - 2 
\sum_{i=1}^n |q_i| = \sum_{i=1}^n \Delta(q_i;\vec{k})
\label{eq:conf_dim_USp2n}
\end{align}
because
 $\Lambda^2([1,0,\ldots,0]) = [0,1,0,\ldots,0] \oplus [0,0,\ldots,0]$
 has non-trivial weights
$  e_i \pm e_j$ , $-(e_i \pm e_j)$ for $1\leq i<j \leq n$
such that $[0,1,0,\ldots,0]$ cancels the vector multiplet contribution 
partially.
In the monopole formula, the magnetic charges $q_i$ are restricted to $q_1\geq 
q_2 \geq \ldots \geq q_n \geq 0$.
Moreover, the dressing factors for a $\usprm(2n)$ gauge node have been 
presented in \cite{Cremonesi:2013lqa}. The shorthand notation $|q_i\pm 
\vec{k}|\equiv \sum_{l=1}^s |q_i\pm k_l|$ summarises the contributions from the 
magnetic charges $k_l$ of the single node in $\bullet$ the $\usprm(2n)$ couples 
to via bifundamental matter.

With this preliminaries at hand, the statement of Proposition \ref{prop:D-type} 
becomes
\begin{align}
\HS_n\equiv \HS_{\MCoulomb \left(T_{\{n\},\bullet} \right)} 
  \stackrel{\text{Prop.\ \ref{prop:D-type}}}{=} 
  \left. \frac{1}{n!} \frac{\diff^n }{\diff \nu^n} \PE \left[\nu \cdot
  \HS_{\MCoulomb \left(T_{\{1\},\bullet} \right)}  \right]  
\right|_{\nu=0} 
\equiv \HS_{\MCoulomb \left(T_{\{1^n\},\bullet} \right) \slash S_n } 
\; . 
\label{eq:HS_explicit_D}
\end{align}
\paragraph{Proof.}
As before, the cycle index \eqref{eq:cycle_index_def} can be employed to 
realise the symmetrisation in \eqref{eq:HS_explicit_D} such that the proof 
proceeds by induction in $n$
\begin{align}
 \HS_n(t) = \frac{1}{n} \sum_{k=1}^n a_k \cdot \HS_{n-k}(t)  \, , \quad a_k 
= f(t^k) \; .
\label{eq:HS_recursive_D}
\end{align}
To begin, one verifies the base case:
\begin{compactenum}[(i)]
 \item $n=1$: trivial, as $\Lambda^2 [1]=0$.
\item $n=2$: The proposal reads
\begin{align}
 \HS_2(t) = \frac{1}{2} \left(a_2 + a_1^2\right)
\end{align}
where the two contributions are treated as follows:
\begin{align}
 a_1^2 &= \sum_{q_1,q_2 \geq 0} P(t,q_1)P(t,q_2) t^{\Delta(q_1)+\Delta(q_2)} 
\notag \\
&=2 \sum_{q_1 > q_2>0} \frac{1}{(1-t)^2} t^{\Delta(q_1)+\Delta(q_2)} 
+2 \sum_{q_1 > 0=q_2} \frac{1}{(1-t)(1-t^2)} t^{\Delta(q_1)+\Delta(0)} \notag
\\
&\quad 
+ \sum_{q_1 = q_2>0} \frac{1}{(1-t)^2} t^{2\Delta(q_1)} 
+\frac{1}{(1-t^2)^2} t^{2\Delta(0)} \; ,\\
a_2 &= \sum_{q\geq 0} P(t^2,q)t^{2\Delta(q)} \notag \\
&= \sum_{q> 0} \frac{1}{1-t^2} t^{2\Delta(q)} + \frac{1}{1-t^4}t^{2\Delta(0)} 
\; .
\end{align}
Adding them up yields
\begin{align}
 \HS_2(t)&= \sum_{q_1 > q_2>0} \frac{1}{(1-t)^2} t^{\Delta(q_1)+\Delta(q_2)} 
 +\sum_{q_1 > 0=q_2} \frac{1}{(1-t)(1-t^2)} t^{\Delta(q_1)+\Delta(0)}  \notag\\
 &\quad +\frac{1}{2} \left( \frac{1}{(1-t)^2}+ \frac{1}{1-t^2} \right) 
  \sum_{q_1 = q_2>0}  t^{2\Delta(q_1)} 
 +\frac{1}{2} \left( \frac{1}{(1-t^2)^2} + \frac{1}{1-t^4} \right) 
t^{2\Delta(0)} \notag \\
&= \sum_{q_1 > q_2>0} \frac{1}{(1-t)^2} t^{\Delta(q_1)+\Delta(q_2)} 
 +\sum_{q_1 > 0=q_2} \frac{1}{(1-t)(1-t^2)} t^{\Delta(q_1)+\Delta(0)}  \\
 &\quad +  \frac{1}{(1-t)(1-t^2)}
  \sum_{q_1 = q_2>0}  t^{2\Delta(q_1)} 
 + \frac{1}{(1-t^2)(1-t^4)} t^{2\Delta(0)} \notag  \; .
\end{align}
Comparing this to the Hilbert series of $\usprm(4)$ with a $\Lambda^2[1,0]$ 
hypermultiplet 
and background charges, one has the conformal dimension 
\eqref{eq:conf_dim_USp2n} and the dressing factors \cite{Cremonesi:2013lqa}
\begin{align}
 P(t,q_1,q_2) &= 
\begin{cases}
 \frac{1}{(1-t)^2} &,\; q_1> q_2>0 \;, \\
 \frac{1}{(1-t)(1-t^2)} &,\; q_1= q_2>0 \;, \\
 \frac{1}{(1-t)(1-t^2)} &,\; q_1> 0=q_2 \;, \\
 \frac{1}{(1-t^2)(1-t^4)} &,\; q_1= q_2=0 \; .
\end{cases}
\end{align}
Consequently, Proposition \ref{prop:D-type} is true for $n=2$.
\end{compactenum}
Next, one proceeds as in the $A$-type case of Section \ref{sec:A-type}, i.e.\ 
the inductive step $(n-1)\to n$. Since there is a slight complication when 
considering $q_1 \geq \ldots\geq q_n  \geq0$, the details of the proof need to 
be elaborated.
\begin{compactenum}[(i)]
 \item $q_1 > \ldots> q_n >0$ can only originate from $a_1 \HS_{n-1}$ 
via
\begin{align}
 \frac{1}{n}a_1 \HS_{n-1} &\supset \frac{1}{n} 
 \sum_{q>0} \frac{1}{1-t} t^{\Delta(q)}  \sum_{q_1> \ldots > q_{n-1} >0} 
\frac{1}{(1-t)^{n_1} } t^{\sum_{i=1}^{n-1}\Delta(q_i)} \notag \\
&\supset 
\frac{1}{(1-t)^{n} }  \sum_{q_1> \ldots > q_{n} >0} 
 t^{\sum_{i=1}^{n}\Delta(q_i)} \; ,
\end{align}
where the $n$ different possibilities to place $q$ between the $(n-1)$ $q_i$ 
eliminated the pre-factor $\frac{1}{n}$.
Moreover, the dressing factor correctly reproduces the 
stabiliser of $q_1 > \ldots> q_n >0$ inside $\usprm(2n)$, i.e.\ $\uo^n$.
\item $q_1 > \ldots > q_{n-l} > 0 = q_{n-l+1} = \ldots =q_{n}$ for which 
contributions arise from $a_1 \HS_{n-1}$ to $a_l \HS_{n-l}$. To start with, 
$a_1 \HS_{n-1}$ provides two contributions
\begin{align}
 a_1 \HS_{n-1} &\supset 
 \sum_{q>0} \frac{1}{1-t} t^{\Delta(q)} \sum_{\substack{l\text{ vanishing } 
q_i \\ \text{out of }n-1} } \frac{1}{(1-t)^{n-l-1} \prod_{a=1}^l (1-t^{2a})} 
t^{\sum_{i=1}^{n-1}\Delta(q_i)} \notag\\
&\supset 
\frac{1}{(1-t)^{n-l}} \frac{n-l}{ \prod_{a=1}^l (1-t^{2a})} 
  \sum_{\substack{l\text{ vanishing } 
q_i \\ \text{out of }n} }
t^{\sum_{i=1}^{n}\Delta(q_i)}
\end{align}
and arranging $q$ between the non-vanishing $q_i$ gives a multiplicity of 
$(n-l)$. The other term is
\begin{align}
 a_1 \HS_{n-1} &\supset 
 \sum_{q=0} \frac{1}{1-t^2} t^{\Delta(0)} \sum_{\substack{(l-1)\text{ vanishing 
} 
q_i \\ \text{out of }n-1} } \frac{1}{(1-t)^{n-l} \prod_{a=1}^{l-1} (1-t^{2a})} 
t^{\sum_{i=1}^{n-1}\Delta(q_i)} \notag\\
  &\supset 
  \frac{1}{(1-t)^{n-l}}  
  \frac{1}{ (1-t^2) \prod_{a=1}^{l-1} (1-t^{2a})}
  \sum_{\substack{l\text{ vanishing } 
q_i \\ \text{out of }n} }  t^{\sum_{i=1}^{n}\Delta(q_i)} \; ,
\end{align}
which has multiplicity one.
Similarly, the contribution form $a_j \HS_{n-j}$ is
\begin{align}
 a_j \HS_{n-j} &\supset \sum_{q>0} \frac{1}{1-t^{2j}} t^{j \Delta(q)} 
 \sum_{ \substack{(l-j)\text{ vanishing 
} q_i \\ \text{out of }n-j}   } \frac{1}{(1-t)^{n-l}} 
\frac{1}{\prod_{a=1}^{l-j} (1-t^{2a})} t^{\sum_{i=1}^{n-j}\Delta(q_i)} \notag\\
&\supset   \frac{1}{(1-t)^{n-l}} 
\frac{1}{(1-t^{2j}) \prod_{a=1}^{l-j} (1-t^{2a})} 
 \sum_{ \substack{l\text{ vanishing 
} q_i \\ \text{out of }n}   } 
t^{\sum_{i=1}^{n}\Delta(q_i)} \; .
\end{align}
Summing up all contributions, one obtains
\begin{align}
 \frac{1}{n} \sum_{j=1}^{l} a_j \HS_{n-j} &\supset 
\frac{1}{n} \frac{1}{(1-t)^{n-l}  \prod_{a=1}^l (1-t^{2a})  } \notag\\
&\qquad \qquad \cdot \left( n-l  + \sum_{m=1}^l \frac{1}{(1-t^{2m})} 
\prod_{b=l-m+1}^{l}(1-t^{2b})  
  \right)
  \sum_{\substack{l\text{ vanishing } 
q_i \\ \text{out of }n} }
t^{\sum_{i=1}^{n}\Delta(q_i)} 
\notag \\
&\supset 
\frac{1}{n} \frac{1}{(1-t)^{n-l}  \prod_{a=1}^l (1-t^{2a})  } 
 \left( n-l  +Q_l(t^2) 
  \right)
  \sum_{\substack{l\text{ vanishing } 
q_i \\ \text{out of }n} }
t^{\sum_{i=1}^{n}\Delta(q_i)} 
\notag\\
&\supset 
\frac{1}{(1-t)^{n-l}  \prod_{a=1}^l (1-t^{2a})  }
  \sum_{\substack{l\text{ vanishing } 
q_i \\ \text{out of }n} }
t^{\sum_{i=1}^{n}\Delta(q_i)}
\end{align}
and one recognises the dressing factor of $\uo^{n-l} \times \usprm(2l)$. Note 
in particular the use of the results of Appendix \ref{app:q-theory}, but this 
time for $Q_l(t^2)=l$.
\item $q_1= \ldots= q_l> q_{l+1} > \ldots > q_{n} > 0$ for which 
contributions arise from $a_1 \HS_{n-1}$ to $a_l \HS_{n-l}$. To start with, 
$a_1 \HS_{n-1}$ provides two contributions
\begin{align}
 a_1 \HS_{n-1} &\supset 
 \sum_{q>0} \frac{1}{1-t} t^{\Delta(q)} \sum_{\substack{l\text{ equal } 
q_i \\ \text{out of }n-1} } \frac{1}{(1-t)^{n-l-1} \prod_{a=1}^l (1-t^{a})} 
t^{\sum_{i=1}^{n-1}\Delta(q_i)} \notag\\
&\supset 
\frac{1}{(1-t)^{n-l}} \frac{n-l}{ \prod_{a=1}^l (1-t^{a})} 
  \sum_{\substack{l\text{ equal } 
q_i \\ \text{out of }n} }
t^{\sum_{i=1}^{n}\Delta(q_i)}
\end{align}
and arranging $q$ between the non-equal $q_i$ gives a multiplicity of 
$n-l$. The other term is
\begin{align}
 a_1 \HS_{n-1} &\supset 
 \sum_{q>0} \frac{1}{1-t} t^{\Delta(q)} \sum_{\substack{(l-1)\text{ equal } 
q_i \\ \text{out of }n-1} } \frac{1}{(1-t)^{n-l} \prod_{a=1}^{l-1} (1-t^{a})} 
t^{\sum_{i=1}^{n-1}\Delta(q_i)} \notag\\
  &\supset 
  \frac{1}{(1-t)^{n-l}}  
  \frac{1}{ (1-t) \prod_{a=1}^{l-1} (1-t^{a})}
  \sum_{\substack{l\text{ equal } 
q_i \\ \text{out of }n} }  t^{\sum_{i=1}^{n}\Delta(q_i)}
\end{align}
which has multiplicity one.
Similarly, the contribution form $a_j \HS_{n-j}$ is
\begin{align}
 a_j \HS_{n-j} &\supset \sum_{q>0} \frac{1}{1-t^{j}} t^{j \Delta(q)} 
 \sum_{ \substack{(l-j)\text{ equal 
} q_i \\ \text{out of }n-j}   } \frac{1}{(1-t)^{n-l}} 
\frac{1}{\prod_{a=1}^{l-j} (1-t^{a})} t^{\sum_{i=1}^{n-j}\Delta(q_i)}  \notag\\
&\supset   \frac{1}{(1-t)^{n-l}} 
\frac{1}{(1-t^{j}) \prod_{a=1}^{l-j} (1-t^{a})} 
 \sum_{ \substack{l\text{ equal 
} q_i \\ \text{out of }n}   } 
t^{\sum_{i=1}^{n}\Delta(q_i)} \; .
\end{align}
Summing up all contributions, one finds
\begin{align}
 \frac{1}{n} \sum_{j=1}^{l} a_j \HS_{n-j} &\supset 
\frac{1}{n} \frac{1}{(1-t)^{n-l}  \prod_{a=1}^l (1-t^{a})  } \notag \\
 &\qquad \qquad \cdot \left( n-l  + \sum_{m=1}^l \frac{1}{(1-t^{m})} 
\prod_{b=l-m+1}^{l}(1-t^{b})  
  \right)
  \sum_{\substack{l\text{ equal } 
q_i \\ \text{out of }n} }
t^{\sum_{i=1}^{n}\Delta(q_i)} \notag \\
&\supset 
\frac{1}{n} \frac{1}{(1-t)^{n-l}  \prod_{a=1}^l (1-t^{a})  } 
 \left( n-l  +Q_l(t) 
  \right)
  \sum_{\substack{l\text{ equal } 
q_i \\ \text{out of }n} }
t^{\sum_{i=1}^{n}\Delta(q_i)} 
\notag\\
&\supset 
\frac{1}{(1-t)^{n-l}  \prod_{a=1}^l (1-t^{a})  }
  \sum_{\substack{l\text{ equal } 
q_i \\ \text{out of }n} }
t^{\sum_{i=1}^{n}\Delta(q_i)}
\end{align}
and one recognises the dressing factor of $\uo^{n-l} \times \urm(l)$.
\item In general, consider a (not necessarily ordered) partition 
$(l_1,\ldots,l_p ; l_0)$ such that $\sum_{j=1}^p l_p +l_0 =n$. Here, $l_0$ 
counts the number of vanishing fluxes, i.e.
\begin{align}
\begin{aligned}
 q_1= \ldots=q_{l_1} &> 
 q_{l_1+1} = \ldots = q_{l_1+l_2} > 
 \ldots >
 q_{l_1+\ldots+l_{p-1}+1 } = \ldots = q_{l_1+\ldots+l_{p} } > 0 \\
 0&= q_{l_1+\ldots+l_{p}+1 } = \ldots = q_{l_1+\ldots+l_{p}+l_0 } \equiv q_n \; 
.
 \end{aligned}
\end{align}
Then from the cases consider above, one obtains
\begin{align}
 \frac{1}{n} \sum_{j=1}^{\max(\{l_j\},l_0)} a_j \HS_{n-j} 
 &\supset
 \frac{1}{n} \frac{1}{\prod_{j=1}^p \prod_{a_j=1}^{l_j}(1-t^{a_j}) \cdot 
\prod_{a_0=1}^{l_0}(1-t^{2a_0})} \notag \\
&\qquad \cdot
\left(n+ \sum_{j=1}^p (Q_{l_j}(t)-l_j)  + (Q_{l_0}(t^2)-l_0) \right) 
\sum_{q's} t^{\sum_{i=1}^{n}\Delta(q_i)} \notag\\
 &\supset
 \frac{1}{\prod_{j=1}^p \prod_{a_j=1}^{l_j}(1-t^{a_j}) \cdot 
\prod_{a_0=1}^{l_0}(1-t^{2a_0})} \sum_{q's} t^{\sum_{i=1}^{n}\Delta(q_i)}
\end{align}
from which one recognises the dressing factor of $\left(\prod_{j=1}^p \urm(l_j) 
\right) \times \usprm(2l_0) $.
\end{compactenum}
Therefore, the pieces together form exactly the Hilbert series for the $n$-th 
step. This concludes the proof of Proposition \ref{prop:D-type}.
\paragraph{Comments.}
The proof establishes that the Coulomb branch of an arbitrary quiver $\bullet$ 
with a $\usprm(2)$-bouquet of size $n$ coincides upon quotient by $S_n$ with 
the Coulomb branch of the quiver $\bullet$ where the bouquet is replaced by a 
$\usprm(2n)$ gauge node with an additional anti-symmetric hypermultiplet.

The nature of the proof allows to draw two immediate corollaries, as in the 
$A$-type case:
\begin{compactenum}[(i)]
 \item One may consider an arbitrary partition $\{n_i\}$ of $n$ such that one 
quotients $T_{\{1^n\},\bullet}$ by $\prod_i S_{n_i}$.
 \item In addition, one may consider quivers with more multiple bouquets, as 
the operation is local on the Coulomb branch.
\end{compactenum}
%
%
\subsection{D-type --- SO(3)-bouquet}
Next, consider a $\sorm(3)$-bouquet in which each $\sorm(3)$-node is equipped 
with a loop corresponding to a hypermultiplet in the 
second symmetric representation. The reason for this will become clear below.
The starting point is again an arbitrary quiver $\bullet$ coupled either to an 
$\sorm(2n+1)$ gauge node with one additional hypermultiplet transforming in 
$\Sym{2}([1,0,\ldots,0])$ or an $\sorm(3)$-bouquet of size $n$. Note that the 
$\sorm(3)$ nodes on the bouquet also have one additional symmetric 
hypermultiplet. Define the following two sets of quivers:
\begin{align}
T_{\{n\},\bullet} =
 \raisebox{-.5\height}{
 \begin{tikzpicture}
	\tikzstyle{gauge} = [circle, draw];
	\tikzstyle{flavour} = [regular polygon,regular polygon sides=4, draw];
	\tikzstyle{extra} = [circle, fill=black, draw];
	\node (e1) [extra] {};
	\node (g1) [gauge,above of=e1,label=right:{$\sorm(2n+1)$}] {};
	\node (g2) [above right of=g1]{$\; \; \Sym{2}$};
	\draw [-] (0.11,1.15) arc (-70:90:10pt);
	\draw [-] (-0.11,1.15) arc (250:90:10pt);
	\draw 	(g1)--(e1);
	\end{tikzpicture}
	}
	\qquad \text{and} \qquad
T_{\{1^n\},\bullet} =
	\raisebox{-.5\height}{
 \begin{tikzpicture}
	\tikzstyle{gauge} = [circle, draw];
	\tikzstyle{flavour} = [regular polygon,regular polygon sides=4, draw];
	\tikzstyle{extra} = [circle, fill=black, draw];
	\node (e1) [extra] {};
	\node (g1) [gauge,above left of =e1,label=left:{$\sorm(3)$}] {};
	\node (g2) [gauge,above right of =e1,label=right:{$\sorm(3)$}] {};
	\node (g3) [above of = e1,label=above:{$n$}] {$\cdots$};
	\draw 	(g1)--(e1)  (g2)--(e1);
	\node (l1) [above left of=g1]{$\Sym{2}$};
	\draw [-] (-0.7+0.11,0.85) arc (-70:90:10pt);
	\draw [-] (-0.7+-0.11,0.85) arc (250:90:10pt);
	\node (l2) [above right of=g2]{$\; \; \Sym{2}$};
	\draw [-] (0.7+0.11,0.85) arc (-70:90:10pt);
	\draw [-] (0.7-0.11,0.85) arc (250:90:10pt);
	\end{tikzpicture}
	} \; .
	\label{eq:SO3-bouquet_quiver}
\end{align}
To clarify, the $\sorm(2n+1)$ as well as all of the $\sorm(3)$ nodes couple to 
the same \emph{single} node in $\bullet$ via bifundamental matter. From the view 
point of the $\sorm(2n+1)$ or $\sorm(3)$ nodes, the quiver 
$\bullet$ contributes background charges $\vec{k}=(k_1,\ldots,k_s)$ for some 
$s\in \NN$, i.e.\ the magnetic charges from the single node they couple to. All 
other contributions from $\bullet$ could be summarised in a function of the 
fugacity, which is not affected by the $S_n$-quotient and is henceforth ignored, 
cf. the discussion below \eqref{eq:A-type_quiver}.
\begin{myProp}
Let $T_{\{n\},\bullet}$ and $T_{\{1^n\},\bullet} $ be as defined in 
\eqref{eq:SO3-bouquet_quiver} then their Coulomb branches satisfy
\begin{align}
\MCoulomb \left(
T_{\{n\},\bullet} 
	\right)
	=
	\MCoulomb \left(
T_{\{1^n\},\bullet} 
	 \right) \slash S_n \;.
\end{align} 
\label{prop:SO3-bouquet}
\end{myProp}
\paragraph{Preliminaries.}
For the proof below, one defines the basic ingredient:
\begin{align}
 f(t)= \HS_{\MCoulomb} \left(
 \raisebox{-.5\height}{
  \begin{tikzpicture}
	\tikzstyle{gauge} = [circle, draw];
	\tikzstyle{flavour} = [regular polygon,regular polygon sides=4, draw];
	\tikzstyle{extra} = [circle, fill=black, draw];
	\node (e1) [extra] {};
	\node (g1) [gauge,above of =e1,label=right:{$\sorm(3)$}] {};
	\node (g2) [above right of=g1]{$\; \; \Sym{2}$};
	\draw [-] (0.11,1.15) arc (-70:90:10pt);
	\draw [-] (-0.11,1.15) arc (250:90:10pt);
	\draw (g1)--(e1);
	\end{tikzpicture} 
	}
 \right) 
 \equiv \HS_{\MCoulomb(T_{\{1\},\bullet})}
\end{align}
which is the Coulomb branch Hilbert series. Here, the conformal dimension reads
\begin{align}
 \Delta(q;\vec{k})=\frac{1}{2}\left(|q-\vec{k}|+|q+\vec{k}|\right) + |q|
\end{align}
for the magnetic charge $q \in \NN$ and background fluxes $\vec{k}$.
The dressing factors associated to $\sorm(3)$ are those of $\usprm(2)$, i.e.
\begin{align}
 P(t,q) = 
 \begin{cases}
  \frac{1}{1-t} & ,\; q>0 \; ,\\ 
  \frac{1}{1-t^2} & ,\; q=0 \; .
 \end{cases}
\end{align}
The $\sorm(2n+1)$ gauge node with one hypermultiplet transforming in 
$\Sym{2}([1,0,\ldots,0])$ and background charges $\vec{k}$ has conformal 
dimension
\begin{align}
 \Delta(q_1,\ldots, q_n;\vec{k}) =
 \frac{1}{2} \sum_{i=1}^n \left( |q_i -\vec{k}| +|q_i +\vec{k}| \right) 
+\sum_{i=1}^n |q_i|
= \sum_{i=1}^n \Delta(q_i;\vec{k}) \; ,
\label{eq:conf_dim_SO2n+1}
\end{align}
because
 $\Sym{2}([1,0,\ldots,0]) = [2,0,\ldots,0] \oplus [0,\ldots,0]$
 with non-trivial weights
  $e_i \pm e_j$, $-(e_i \pm e_j)$ for $1\leq i<j \leq n$ 
 and $\pm 2e_i$ for $1\leq i \leq n$
such that $[2,0,\ldots,0]$ cancels the vector multiplet contribution. 
In the monopole formula, the magnetic charges $q_i$ are restricted to $q_1 
\geq q_2 \geq \ldots \geq q_n \geq 0$.
Moreover, the dressing factors of $\sorm(2n+1)$ are those of $\usprm(2n)$, see 
\cite{Cremonesi:2013lqa}. The shorthand notation $|q_i\pm 
\vec{k}|\equiv \sum_{l=1}^s |q_i\pm k_l|$ summarises the contributions from the 
magnetic charges $k_l$ of the single node in $\bullet$ the $\sorm(2n+1)$ 
couples to via bifundamental matter.
\paragraph{Proof.}
To prove Proposition \ref{prop:SO3-bouquet}, one needs to verify the Hilbert 
series relations \eqref{eq:HS_explicit_A} or \eqref{eq:HS_explicit_D} for the 
case of a $\sorm(3)$-bouquet. 
As before, the proof relies on the recursive formula 
\eqref{eq:cycle_index_recursive} of the cycle index and proceeds by 
induction as in \eqref{eq:HS_recursive_D}.
As a first step, one considers the base case.
\begin{compactenum}[(i)]
 \item $n=1$: trivial.
\item $n=2$: The proposal reads
\begin{align}
 \HS_2(t) = \frac{1}{2} \left(a_2 + a_1^2\right) \qquad \text{with} \quad a_k 
\coloneqq f(t^k) \; ,
\end{align}
where the two contributions are treated as follows:
\begin{align}
 a_1^2 &= \sum_{q_1,q_2 \geq 0} P(t,q_1)P(t,q_2) t^{\Delta(q_1)+\Delta(q_2)} 
\notag \\
&=2 \sum_{q_1 > q_2>0} \frac{1}{(1-t)^2} t^{\Delta(q_1)+\Delta(q_2)} 
+2 \sum_{q_1 > 0=q_2} \frac{1}{(1-t)(1-t^2)} t^{\Delta(q_1)+\Delta(0)} \notag
\\
&\quad 
+ \sum_{q_1 = q_2>0} \frac{1}{(1-t)^2} t^{2\Delta(q_1)} 
+\frac{1}{(1-t^2)^2} t^{2\Delta(0)} \; ,\\
a_2 &= \sum_{q\geq 0} P(t^2,q)t^{2\Delta(q)} \notag \\
&= \sum_{q> 0} \frac{1}{1-t^2} t^{2\Delta(q)} + \frac{1}{1-t^4}t^{2\Delta(0)} 
\; .
\end{align}
Adding them up, yields
\begin{align}
 \HS_2(t)&= \sum_{q_1 > q_2>0} \frac{1}{(1-t)^2} t^{\Delta(q_1)+\Delta(q_2)} 
 +\sum_{q_1 > 0=q_2} \frac{1}{(1-t)(1-t^2)} t^{\Delta(q_1)+\Delta(0)}  \notag \\
 &\quad +\frac{1}{2} \left( \frac{1}{(1-t)^2}+ \frac{1}{1-t^2} \right) 
  \sum_{q_1 = q_2>0}  t^{2\Delta(q_1)} 
 +\frac{1}{2} \left( \frac{1}{(1-t^2)^2} + \frac{1}{1-t^4} \right) 
t^{2\Delta(0)} \notag \\
&= \sum_{q_1 > q_2>0} \frac{1}{(1-t)^2} t^{\Delta(q_1)+\Delta(q_2)} 
 +\sum_{q_1 > 0=q_2} \frac{1}{(1-t)(1-t^2)} t^{\Delta(q_1)+\Delta(0)}  \\
 &\quad +  \frac{1}{(1-t)(1-t^2)}
  \sum_{q_1 = q_2>0}  t^{2\Delta(q_1)} 
 + \frac{1}{(1-t^2)(1-t^4)} t^{2\Delta(0)}  \; .\notag 
\end{align}
Comparing this to the monopole formula of $\sorm(5)$ with one $\Sym{2}[1,0]$ 
hypermultiplet and background charges, the conformal dimension follows from 
\eqref{eq:conf_dim_SO2n+1} and the dressing factors read 
\begin{align}
 P(t,q_1,q_2) = 
\begin{cases}
 \frac{1}{(1-t)^2} &\; q_1> q_2>0 \;, \\
 \frac{1}{(1-t)(1-t^2)} &\; q_1= q_2>0 \;, \\
 \frac{1}{(1-t)(1-t^2)} &\; q_1> 0=q_2 \;, \\
 \frac{1}{(1-t^2)(1-t^4)} &\; q_1= q_2=0 \; .
\end{cases}
\end{align}
Consequently, Proposition \ref{prop:SO3-bouquet} is true for $n=2$.
\end{compactenum}
The argument proceeds as in the $D$-type case of Section \ref{sec:D-type}, one 
proves the inductive step $(n-1)\to n$. As the dressing factors as well as 
the lattice of magnetic charges are identical to the $\usprm(2n)$ case, it is 
unnecessary to spell out the details of the proof. The only 
point to appreciate is that the conformal dimension is the sum of the 
individual $\sorm(3)$ conformal dimensions.
\paragraph{Comments.}
Again, the same corollaries are in order: (i) one can generalise to arbitrary 
partitions, and (ii) one can consider multiple bouquets.

Moreover, note that the Coulomb branch of 
\begin{align}
 T_{1,n}^{\text{D}'\text{-type}}= 
 \raisebox{-.5\height}{
  \begin{tikzpicture}
	\tikzstyle{gauge} = [circle, draw];
	\tikzstyle{flavour} = [regular polygon,regular polygon sides=4, draw];
	\tikzstyle{extra} = [circle, fill=black, draw];
	\node (f1) [flavour,label=right:{$\usprm(2n)$}] {};
	\node (g1) [gauge,above of =f1,label=right:{$\sorm(3)$}] {};
	\node (g2) [above right of=g1]{$\; \; \Sym{2}$};
	\draw [-] (0.11,1.15) arc (-70:90:10pt);
	\draw [-] (-0.11,1.15) arc (250:90:10pt);
	\draw (g1)--(f1);
	\end{tikzpicture} 
	}
	\qquad \text{with } \qquad
	\HS_{T_{1,n}^{\text{D}'\text{-type}}} = 
\frac{1-t^{2n+4}}{(1-t^2)(1-t^{n+1})(1-t^{n+2})} 
\end{align}
is the $D_{n+3}$-singularity. Hence, Proposition \ref{prop:SO3-bouquet} 
provides the missing analogue of \eqref{eq:Ex_U(k)}, \eqref{eq:Ex_Sp(k)} for 
\eqref{eq:Dn+3-type}, i.e.
\begin{align}
 \MCoulomb \left(
 \raisebox{-.5\height}{
  \begin{tikzpicture}
	\tikzstyle{gauge} = [circle, draw];
	\tikzstyle{flavour} = [regular polygon,regular polygon sides=4, draw];
	\tikzstyle{extra} = [circle, fill=black, draw];
	\node (f1) [flavour,label=right:{$\usprm(2n)$}] {};
	\node (g1) [gauge,above of =f1,label=right:{$\sorm(2k+1)$}] {};
	\node (g2) [above right of=g1]{$\; \; \Sym{2}$};
	\draw [-] (0.11,1.15) arc (-70:90:10pt);
	\draw [-] (-0.11,1.15) arc (250:90:10pt);
	\draw (g1)--(f1);
	\end{tikzpicture} 
	}
 \right) =
 \Sym{k} \left( 
 \MCoulomb \left(
 \raisebox{-.5\height}{
  \begin{tikzpicture}
	\tikzstyle{gauge} = [circle, draw];
	\tikzstyle{flavour} = [regular polygon,regular polygon sides=4, draw];
	\tikzstyle{extra} = [circle, fill=black, draw];
	\node (f1) [flavour,label=right:{$\usprm(2n)$}] {};
	\node (g1) [gauge,above of =f1,label=right:{$\sorm(3)$}] {};
	\node (g2) [above right of=g1]{$\; \; \Sym{2}$};
	\draw [-] (0.11,1.15) arc (-70:90:10pt);
	\draw [-] (-0.11,1.15) arc (250:90:10pt);
	\draw (g1)--(f1);
	\end{tikzpicture} 
	}
 \right) \right) \; .
\end{align}
%
%
\section{Other applications}
\label{sec:applications}
After establishing the generalisations underlying the $A$ and $D$-type 
singularities, one may wonder if there are other types of bouquets that can be 
considered.
One could ask what are sufficient conditions such that a $G_n$ gauge node, 
which may be supplemented by additional matter, can be obtained from 
an $S_n$-quotient of a certain $G_1$-bouquet. To be precise, the $G_n$ node 
as well as all nodes of the $G_1$-bouquet are coupled to the same \emph{single} 
node in a given quiver gauge theory via bifundamental matter. From the 
aforementioned cases one formulates three conditions: 
\begin{compactenum}[(i)]
 \item Additivity of the conformal dimension: i.e.\ $\Delta$ of $G_n$ is the 
sum of the conformal dimensions of the $G_1$ nodes.
\item Compatibility of the GNO lattices.
\item Compatibility of the dressing factors.
\end{compactenum}
While the first statement is concise, the second and third are less precise. 
However, by recalling \cite{Hanany:2016ezz,Hanany:2016pfm} the interpretation 
of the dressing factors $P(t,q_i)$ as Hilbert series of $\C[\gfrak]^G \cong 
\C[\tfrak]^{\Wcal_G}$, with $\tfrak$ a Cartan sub-algebra of $\gfrak$ and 
$\Wcal_G$ the Weyl group, one can identify all classical groups that allow 
for a $S_n$ factor in $\Wcal_G$.

Suppose $\Wcal_{G_n} = S_n \ltimes (\Gamma)^n$ and denote the chosen Cartan 
sub-algebra as $\tfrak \cong V^n$, for some $1$-dimensional vector space $V$, 
then
\begin{align}
 \C[\Lie(G_n)]^{G_n} \cong \C[V^{n}]^{S_n \ltimes (\Gamma)^n} 
 \cong \Sym{n}\left(\C[V]^{\Gamma}  \right)
 \cong \Sym{n}\left(\C[\Lie(G_1)]^{G_1}  \right)
\end{align}
and a similar argument is valid for the summation ranges in the monopole 
formula of $G_n$ and $G_1$.
\begin{table}
 \centering
 \begin{tabular}{c|c|c}
 \toprule
  $\gfrak$ & $\Wcal$ & adjoint of $G$  \\\midrule
$A_n$ & $S_{n+1}$ & $[1,0,\ldots,0,1]$ \\
$B_n$ & $S_n \ltimes (\Z_2)^n$ & $\Lambda^2([1,0,\ldots,0])$ \\
$C_n$ & $S_n \ltimes (\Z_2)^n$ & $\Sym{2}([1,0,\ldots,0])$ \\
$D_n$ & $S_n \ltimes (\Z_2)^{n-1}$ & $\Lambda^{2}([1,0,\ldots,0])$ \\ 
\bottomrule
 \end{tabular}
\caption{Classical algebras and their Weyl groups.}
\label{tab:Weyl_groups}
\end{table}
By inspecting classical Weyl groups in Table \ref{tab:Weyl_groups} one 
concludes that choosing $G_n$ to be either $\urm(n)$, $\sorm(2n+1)$, 
$\usprm(2n)$, or $\orm(2n)$ together with one adjoint hypermultiplet leads to 
possible $S_n$-quotients on the Coulomb branch. All the different cases are 
elaborated on in the subsequent sections. The only exception is $\urm(n)$ as it 
agrees with Proposition \ref{prop:A-type}.
%
%
\subsection{SO(3)-bouquet}
Specifying the above to a $\sorm(2n+1)$ gauge node with one adjoint 
hypermultiplet coupled to an arbitrary quiver $\bullet$, one finds:
\begin{myCor}
Let $T_{\{n\},\bullet}$ and $T_{\{1^n\},\bullet} $ be defined as
\begin{align}
T_{\{n\},\bullet} =
 \raisebox{-.5\height}{
 \begin{tikzpicture}
	\tikzstyle{gauge} = [circle, draw];
	\tikzstyle{flavour} = [regular polygon,regular polygon sides=4, draw];
	\tikzstyle{extra} = [circle, fill=black, draw];
	\node (e1) [extra] {};
	\node (g1) [gauge,above of=e1,label=right:{$\sorm(2n+1)$}] {};
	\node (g2) [above right of=g1]{$\Adj$};
	\draw [-] (0.11,1.15) arc (-70:90:10pt);
	\draw [-] (-0.11,1.15) arc (250:90:10pt);
	\draw 	(g1)--(e1);
	\end{tikzpicture}
	}
	\qquad \text{and} \qquad
T_{\{1^n\},\bullet} =
	\raisebox{-.5\height}{
 \begin{tikzpicture}
	\tikzstyle{gauge} = [circle, draw];
	\tikzstyle{flavour} = [regular polygon,regular polygon sides=4, draw];
	\tikzstyle{extra} = [circle, fill=black, draw];
	\node (e1) [extra] {};
	\node (g1) [gauge,above left of =e1,label=left:{$\sorm(3)$}] {};
	\node (g2) [gauge,above right of =e1,label=right:{$\sorm(3)$}] {};
	\node (g3) [above of = e1,label=above:{$n$}] {$\cdots$};
		\draw 	(g1)--(e1)  (g2)--(e1);
	\node (l1) [above left of=g1]{$\Adj$};
	\draw [-] (-0.7+0.11,0.85) arc (-70:90:10pt);
	\draw [-] (-0.7+-0.11,0.85) arc (250:90:10pt);
	\node (l2) [above right of=g2]{$\Adj$};
	\draw [-] (0.7+0.11,0.85) arc (-70:90:10pt);
	\draw [-] (0.7-0.11,0.85) arc (250:90:10pt);
	\end{tikzpicture}
	} \; ,
	\label{eq:SO3-bouquet_adj}
\end{align}
then their Coulomb branches satisfy 
\begin{align}
\MCoulomb \left(
T_{\{n\},\bullet} 
	\right)
	=
	\MCoulomb \left(
T_{\{1^n\},\bullet} 
	 \right) \slash S_n \;.
\end{align} 
\label{cor:SO3-bouquet_adj}
\end{myCor}
To prove Corollary \ref{cor:SO3-bouquet_adj} one follows all the steps of the 
proof of Proposition \ref{prop:SO3-bouquet}. The only point the take care of is 
that the conformal dimensions add up, which is not difficult to see.

\subsection{USp(2)-bouquet}
Next, consider a $\usprm(2n)$ gauge node with one 
adjoint hypermultiplet coupled to an arbitrary quiver $\bullet$ via 
bifundamental hypermultiplets.
\begin{myCor}
Let $T_{\{n\},\bullet}$ and $T_{\{1^n\},\bullet} $ be defined as
\begin{align}
T_{\{n\},\bullet} =
 \raisebox{-.5\height}{
 \begin{tikzpicture}
	\tikzstyle{gauge} = [circle, draw];
	\tikzstyle{flavour} = [regular polygon,regular polygon sides=4, draw];
	\tikzstyle{extra} = [circle, fill=black, draw];
	\node (e1) [extra] {};
	\node (g1) [gauge,above of=e1,label=right:{$\usprm(2n)$}] {};
	\node (g2) [above right of=g1]{$\Adj$};
	\draw [-] (0.11,1.15) arc (-70:90:10pt);
	\draw [-] (-0.11,1.15) arc (250:90:10pt);
	\draw 	(g1)--(e1);
	\end{tikzpicture}
	}
	\qquad \text{and} \qquad
T_{\{1^n\},\bullet} =
	\raisebox{-.5\height}{
 \begin{tikzpicture}
	\tikzstyle{gauge} = [circle, draw];
	\tikzstyle{flavour} = [regular polygon,regular polygon sides=4, draw];
	\tikzstyle{extra} = [circle, fill=black, draw];
	\node (e1) [extra] {};
	\node (g1) [gauge,above left of =e1,label=left:{$\usprm(2)$}] {};
	\node (g2) [gauge,above right of =e1,label=right:{$\usprm(2)$}] {};
	\node (g3) [above of = e1,label=above:{$n$}] {$\cdots$};
		\draw 	(g1)--(e1)  (g2)--(e1);
	\node (l1) [above left of=g1]{$\Adj$};
	\draw [-] (-0.7+0.11,0.85) arc (-70:90:10pt);
	\draw [-] (-0.7+-0.11,0.85) arc (250:90:10pt);
	\node (l2) [above right of=g2]{$\Adj$};
	\draw [-] (0.7+0.11,0.85) arc (-70:90:10pt);
	\draw [-] (0.7-0.11,0.85) arc (250:90:10pt);
	\end{tikzpicture}
	} \; ,
	\label{eq:USp2-bouquet_adj}
\end{align}
then their Coulomb branches satisfy 
\begin{align}
\MCoulomb \left(
T_{\{n\},\bullet} 
	\right)
	=
	\MCoulomb \left(
T_{\{1^n\},\bullet} 
	 \right) \slash S_n \;.
\end{align} 
\label{cor:USp2-bouquet_adj}
\end{myCor}
The proof of Corollary \ref{cor:USp2-bouquet_adj} follows from the 
proof of Proposition \ref{prop:D-type}, by verifying that the conformal 
dimensions add up appropriately.
\subsection{O(2)-bouquet}
Finally, let an arbitrary quiver $\bullet$ be coupled either 
to an $\orm(2n)$ gauge node with one additional anti-symmetric hypermultiplet 
or to a $\orm(2)$-bouquet of size $n$. 
\begin{myCor}
For the quiver gauge theories $T_{\{n\},\bullet}$ and $T_{\{1^n\},\bullet}$, 
defined as  
\begin{align}
T_{\{n\},\bullet} =
 \raisebox{-.5\height}{
 \begin{tikzpicture}
	\tikzstyle{gauge} = [circle, draw];
	\tikzstyle{flavour} = [regular polygon,regular polygon sides=4, draw];
	\tikzstyle{extra} = [circle, fill=black, draw];
	\node (e1) [extra] {};
	\node (g1) [gauge,above of=e1,label=right:{$\orm(2n)$}] {};
	\node (g2) [above right of=g1]{$\Lambda^2$};
	\draw [-] (0.11,1.15) arc (-70:90:10pt);
	\draw [-] (-0.11,1.15) arc (250:90:10pt);
	\draw 	(g1)--(e1);
	\end{tikzpicture}
	}
 \qquad \text{and} \qquad 
 T_{\{1^n\},\bullet} =
 \raisebox{-.5\height}{
 \begin{tikzpicture}
	\tikzstyle{gauge} = [circle, draw];
	\tikzstyle{flavour} = [regular polygon,regular polygon sides=4, draw];
	\tikzstyle{extra} = [circle, fill=black, draw];
	\node (e1) [extra] {};
	\node (g1) [gauge,above left of =e1,label=above:{$\orm(2)$}] {};
	\node (g2) [gauge,above right of =e1,label=above:{$\orm(2)$}] {};
	\node (g3) [above of = e1,label=above:{$n$}] {$\cdots$};
	\draw (g1)--(e1)  (g2)--(e1);
	\end{tikzpicture}
	} 
	\label{eq:O2-bouquet_quiver}
\end{align}
the Coulomb branches satisfy
\begin{align}
\MCoulomb \left(T_{\{n\},\bullet} \right) =
\MCoulomb \left(T_{\{1^n\},\bullet} \right) \slash S_n \; .
\end{align} 
\label{cor:O2-bouquet}
\end{myCor}
Since this is the first time $O(2n)$ gauge nodes appear, some remarks on the 
proof are in order. Analogous to Section \ref{sec:A_and_D-type}, define the 
basic ingredient:
\begin{align}
 \HS_{\MCoulomb} \left(
 \raisebox{-.5\height}{
  \begin{tikzpicture}
	\tikzstyle{gauge} = [circle, draw];
	\tikzstyle{flavour} = [regular polygon,regular polygon sides=4, draw];
	\tikzstyle{extra} = [circle, fill=black, draw];
	\node (e1) [extra] {};
	\node (g1) [gauge,above of =e1,label=above:{$\orm(2)$}] {};
	\draw 	(g1)--(e1)  ;
	\end{tikzpicture} 
	}
 \right) 
 \equiv
 \HS_{\MCoulomb(T_{\{1\},\bullet})}
 \label{eq:O2-Hilbert-series}
\end{align}
which is the Coulomb branch Hilbert series. Here, the conformal dimension reads
\begin{align}
 \Delta(q;\vec{k})=\frac{1}{2}(|q-\vec{k}|+|q+\vec{k}|) 
\end{align}
for the magnetic charge $q \in \NN$ and background fluxes $\vec{k}$.
Following \cite{Cremonesi:2014uva}, the dressing factors associated to $\orm(2)$ 
are those of $\sorm(3)$, i.e.
\begin{align}
 P(t,q) = 
 \begin{cases}
  \frac{1}{1-t} & ,\; q>0 \; ,\\ 
  \frac{1}{1-t^2} & ,\; q=0 \; .
 \end{cases}
\end{align}
Note that there is no extra topological fugacity for $\orm(2)$.

The $\orm(2n)$ gauge node with one hypermultiplet 
transforming as $\Lambda^2([1,0,\ldots,0])$ and background charges has 
conformal dimension
\begin{align}
 \Delta(q_1,\ldots, q_n;\vec{k}) = 
 \frac{1}{2} \sum_{i=1}^n \left( |q_i -\vec{k}| + |q_i +\vec{k}| \right)  = 
\sum_{i=1}^n \Delta(q_i;\vec{k})
\label{eq:conf_dim_O2n}
\end{align}
because $\Lambda^2([1,0,\ldots,0]) = [0,1,0,\ldots,0]$ with non-trivial weights
$e_i \pm e_j$, $-(e_i \pm e_j)$ for $1\leq i<j \leq n$ such that 
$[0,1,0,\ldots,0]$ cancels the vector multiplet contribution partially. Again, 
the magnetic charges $q_i$ satisfy $q_1\geq q_2 \geq \ldots \geq q_n\geq 0$ in 
the monopole formula. The dressing factors for $\orm(2n)$ have been discussed 
in \cite{Cremonesi:2014uva}. The shorthand notation $|q_i\pm 
\vec{k}|\equiv \sum_{l=1}^s |q_i\pm k_l|$ summarises the contributions from the 
magnetic charges $k_l$ of the single node in $\bullet$ the $\orm(2n)$ couples 
to via bifundamental matter.
 
As the dressing factors and GNO lattice for $\orm(2n)$ originate from 
$\sorm(2n+1)$, which are the same as for $\usprm(2n)$, the proof of Corollary 
\ref{cor:O2-bouquet} is consequence of the proofs of Propositions 
\ref{prop:D-type} and \ref{prop:SO3-bouquet}.
%
%
\subsection{Remarks and example}
With Corollaries \ref{cor:SO3-bouquet_adj}--\ref{cor:O2-bouquet} at ones 
disposal, one can immediately generalise to the following:
\begin{myCor}
 Let $G_n$ be either $\urm(n)$, $\sorm(2n+1)$, $\usprm(2n)$, or $\orm(2n)$ and 
$\{n_i\}$ be a partition of $n$. The Coulomb branch of the quiver gauge theory
\begin{align}
T_{\{n_i\},\bullet} =
 \raisebox{-.5\height}{
 \begin{tikzpicture}
	\tikzstyle{gauge} = [circle, draw];
	\tikzstyle{flavour} = [regular polygon,regular polygon sides=4, draw];
	\tikzstyle{extra} = [circle, fill=black, draw];
	\node (e1) [extra] {};
	\node (g1) [gauge,above left of =e1,label=left:{$G_{n_1}$}] {};
	\node (g2) [gauge,above right of =e1,label=right:{$G_{n_l}$}] {};
	\node (g3) [above of = e1] {$\cdots$};
	\draw 	(g1)--(e1)  (g2)--(e1);
	\node (l1) [above left of=g1]{$\Adj$};
	\draw [-] (-0.7+0.11,0.85) arc (-70:90:10pt);
	\draw [-] (-0.7+-0.11,0.85) arc (250:90:10pt);
	\node (l2) [above right of=g2]{$\Adj$};
	\draw [-] (0.7+0.11,0.85) arc (-70:90:10pt);
	\draw [-] (0.7-0.11,0.85) arc (250:90:10pt);
	\end{tikzpicture}
	}
\end{align}
satisfies
\begin{align}
 \MCoulomb\left(T_{\{n_i\},\bullet} \right) = 
\MCoulomb\left(T_{\{1^n\},\bullet} \right) \slash\prod_i S_{n_i} \; .
\end{align}
\label{cor:full_quotient}
\end{myCor}
Likewise, one could consider quiver gauge theories coupled to various bouquets 
at different nodes. 
\paragraph{Example.}
Before closing, it is interesting to study an example of Corollary 
\ref{cor:O2-bouquet}. This highlights the use of the monopole formula as 
very fortunate because the corresponding statements on rational functions would 
have been very cumbersome to prove.
To begin with, one readily computes \eqref{eq:O2-Hilbert-series} for $\bullet$ 
being a flavour node and obtains
\begin{align}
\HS_{\MCoulomb(T_{\{1\},\Box}) 
}(t)=\frac{1-t^{2k+2}}{(1-t^2)(1-t^k)(1-t^{k+1})} 
 \qquad \text{for}\quad
 T_{\{1\},\Box}=
 \raisebox{-.5\height}{
  \begin{tikzpicture}
	\tikzstyle{gauge} = [circle, draw];
	\tikzstyle{flavour} = [regular polygon,regular polygon sides=4, draw];
	\node (f1) [flavour,label=right:{$\usprm(2k)$}] {};
	\node (g1) [gauge,above of =f1,label=right:{$\orm(2)$}] {};
	\draw 	(g1)--(f1)  ;
	\end{tikzpicture} 
	}  \; .
\end{align}
A similar computation can be performed for $\orm(6)$ with one adjoint 
hypermultiplet and $k$ flavours. In detail:
\begin{align}
&\HS_{\MCoulomb(T_{\{3\},\Box}) 
}(t)=\frac{f(t)}{g(t)}
 \qquad \text{for}\qquad
 T_{\{3\},\Box}=
 \raisebox{-.5\height}{
  \begin{tikzpicture}
	\tikzstyle{gauge} = [circle, draw];
	\tikzstyle{flavour} = [regular polygon,regular polygon sides=4, draw];
	\node (f1) [flavour,label=right:{$\usprm(2k)$}] {};
	\node (g1) [gauge,above of =f1,label=right:{$\orm(6)$}] {};
	\draw 	(g1)--(f1)  ;
	\node (g2) [above right of=g1]{$\Lambda^2$};
	\draw [-] (0.11,1.15) arc (-70:90:10pt);
	\draw [-] (-0.11,1.15) arc (250:90:10pt);
	\end{tikzpicture} 
	}  \; ,\\
 f(t)&=1+
 t^{ k+1} (1+t+t^2+t^3+t^4)
 +t^{2 k+1} (1 +2t +2t^2 +2t^3 +2t^4 +t^5 +t^6)
 \notag\\
 &\quad 
 +t^{3 k+1} (1 +t +3t^2 +2t^3 +2t^4 +3t^5 +t^6 +t^7)
 \notag\\
 &\quad 
 +t^{4 k+2} (1 +t +2t^2 +2t^3 +2t^4 +2t^5 +t^6)
 +t^{5 k+4} (1+t+t^2+t^3+t^4 ) + t^{6k+9}
 \; ,\notag\\
 g(t)&=
 \left(1-t^2\right) 
 \left(1-t^4\right) 
 \left(1-t^{6}\right) 
 \left(1-t^{ k}\right) 
 \left(1-t^{2 k}\right) 
 \left(1-t^{3 k}\right) \notag \;.
\end{align}
Then Corollary \ref{cor:O2-bouquet} becomes equivalent to the claim
\begin{align}
\begin{aligned}
 \HS_{\MCoulomb(T_{\{3\},\Box})} (t) 
 =
 \frac{1}{3!} 
 \bigg( 
 \left(\HS_{\MCoulomb(T_{\{1\},\Box})}(t)\right)^3
 &+3 \cdot \HS_{\MCoulomb(T_{\{1\},\Box})}(t^2)\cdot 
\HS_{\MCoulomb(T_{\{1\},\Box})}(t) \\
 &+2 \cdot \HS_{\MCoulomb(T_{\{1\},\Box})}(t^3)
 \bigg) \; ,
 \end{aligned}
\end{align}
which can be verified explicitly by inserting the rational functions.
%
%
\section{Discussion and conclusions}
\label{sec:conclusion}
In this note we have shown that discrete $S_n$-quotients on Coulomb branches of 
quivers with various bouquets are entirely local operations. By this we mean 
that the geometric $S_n$-quotient on $\MCoulomb$ is realised on the quiver (and 
the monopole formula) as an operation on the bouquet alone; the remainder of 
the quiver is untouched by the $S_n$ action.

We provided the $A$ and $D$-type Propositions 
\ref{prop:A-type}--\ref{prop:SO3-bouquet} in Section \ref{sec:A_and_D-type} 
and proved them via the cycle index for $S_n$. 
Subsequently, we explore various other possibilities in Section 
\ref{sec:applications} and derived Corollaries 
\ref{cor:SO3-bouquet_adj}--\ref{cor:full_quotient}.  In comparison, the gauge 
nodes in Section \ref{sec:A_and_D-type} are supplemented by loops corresponding 
to matter as in the ADHM quivers, whereas the gauge nodes in Section 
\ref{sec:applications} are equipped with one additional adjoint hypermultiplet. 
The $A$-type case of $\urm(n)$ nodes is the only scenario for which both 
notions coincide.

The results are important for a number of reasons: firstly, it allows to deduce 
if certain $3$-dimensional $\Ncal=4$ Coulomb branches are $S_n$ orbifolds of 
one another. For instance, the sub-regular nilpotent orbit of $G_2$ is an $S_3$ 
quotient of the minimal nilpotent orbit of $\sorm(8)$, cf. 
\cite{Brylinski:1992}. Due to the discrete 
quotient proposition, the statement follows immediately by inspecting the
$3$-dimensional $\Ncal=4$ quivers
\begin{align}
 T_{\{1^3\},\Box\text{---}\circ}&= 
 \raisebox{-.5\height}{
 \begin{tikzpicture}
	\tikzstyle{gauge} = [circle, draw];
	\tikzstyle{flavour} = [regular polygon,regular polygon sides=4, draw];
	\tikzstyle{extra} = [circle, fill=black, draw];
	\node (f1) [flavour,label=below:{$\uo$}] {};
	\node (g1) [gauge,right of=f1,label=below:{$\urm(2)$}] {};
	\node (g2) [gauge,above left of=g1,label=above left:{$\urm(1)$}] {};
	\node (g3) [gauge,above of=g1,label=above:{$\urm(1)$}] {};
	\node (g4) [gauge,above right of=g1,label=above right:{$\urm(1)$}] {};
	\draw 	(g1)--(f1) (g1)--(g2) (g1)--(g3) (g1)--(g4);
	\end{tikzpicture}
	}	
\quad \xrightarrow[\text{quotient}]{ S_3 } \quad
T_{\{3\},\Box\text{---}\circ}=
\raisebox{-.5\height}{
 \begin{tikzpicture}
	\tikzstyle{gauge} = [circle, draw];
	\tikzstyle{flavour} = [regular polygon,regular polygon sides=4, draw];
	\tikzstyle{extra} = [circle, fill=black, draw];
	\node (f1) [flavour,label=below:{$\uo$}] {};
	\node (g1) [gauge,right of=f1,label=below:{$\urm(2)$}] {};
	\node (g2) [gauge,above of=g1,label=right:{$\urm(3)$}] {};
	\node (g3) [above right of=g2]{$\Adj$};
	\draw [-] (1+0.11,1.15) arc (-70:90:10pt);
	\draw [-] (1-0.11,1.15) arc (250:90:10pt);
	\draw 	(g1)--(f1) (g1)--(g2);
	\end{tikzpicture}
	}	
  \\
\clorbit{\text{subreg}}^{G_2} &= 
\MCoulomb \left( T_{\{3\},\Box\text{---}\circ}\right)
=
\MCoulomb\left(T_{\{1^3\},\Box\text{---}\circ} 
\right) \slash S_3
= \clorbit{\text{min}}^{\sorm(8)} \slash S_3 \;.
\end{align}
Secondly, the propositions allow to systematically study the different phases 
of $6$-dimensional Higgs branches as put forward by \cite{Hanany:2018vph}. 
For instance, Proposition \ref{prop:D-type} allows to conclude a similar 
statement to \eqref{eq:Ex_6d_2} on the different phases of the Higgs branches of 
multiple M5-branes on a $\C^2 \slash D_k$ singularity \cite{Hanany:2018uhm}. 
The conjecture becomes that for a partition $\{n_i\}$ of $n$, such that the 
M5-branes coincide in a pattern of $n_i$, the $3$-dimensional quiver reads 
\begin{align}
F_{n,k}^{\mathrm{D}}=
  \raisebox{-.5\height}{
 	\begin{tikzpicture}
	\tikzstyle{gauge} = [circle, draw];
	\tikzstyle{flavour} = [regular polygon,regular polygon sides=4, draw];
\node (g1) [gauge,label={[rotate=-45]below right:$\orm(2)$}] {};
\node (g2) [gauge,right of=g1,label={[rotate=-45]below right:$\usprm(2)$}] {};
\node (g3) [gauge,right of=g2,label={[rotate=-45]below right:$\orm(4)$}] {};
\node (g4) [gauge,right of=g3,label={[rotate=-45]below right:$\usprm(4)$}] {};
\node (g5) [right of=g4] {$\ldots$};
\node (g6) [gauge,right of=g5,label={[rotate=-45]below right:$\orm(2k{-}2)$}] 
{};
\node (g7) [gauge,right of=g6,label={[rotate=-45]below right:$\usprm(2k{-}2)$}] 
{};
\node (g8) [gauge,right of=g7,label={[rotate=-45]below right:$\orm(2k)$}] {};
\node (g9) [gauge,right of=g8,label={[rotate=-45]below right:$\usprm(2k{-}2)$}] 
{};
\node (g10) [gauge,right of=g9,label={[rotate=-45]below right:$\orm(2k{-}2)$}] 
{};
\node (g11) [right of=g10] {$\ldots$};
\node (g12) [gauge,right of=g11,label={[rotate=-45]below 
right:$\usprm(4)$}] {};
\node (g13) [gauge,right of=g12,label={[rotate=-45]below right:$\orm(4)$}] 
{};
\node (g14) [gauge,right of=g13,label={[rotate=-45]below 
right:$\usprm(2)$}] {};
\node (g15) [gauge,right of=g14,label={[rotate=-45]below right:$\orm(2)$}] 
{};
\node (g16) [gauge,above left of=g8,label=left:{$\usprm(2n_1)$}] {};
\node (g17) [above of=g8] {$\ldots$};
\node (g18) [gauge,above right of=g8,label=right:{$\usprm(2n_l)$}] {};
	\draw [-] (6.3+0.11,0.85) arc (-70:90:10pt);
	\draw [-] (6.3-0.11,0.85) arc (250:90:10pt);
	\node (l1) [above left of=g16] {$\Lambda^2$};
	\draw [-] (7.7+0.11,0.85) arc (-70:90:10pt);
	\draw [-] (7.7-0.11,0.85) arc (250:90:10pt);
	\node (l2) [,above right of=g18] {$\Lambda^2$};
	\draw (g1)--(g2) (g2)--(g3) (g3)--(g4) (g4)--(g5) (g5)--(g6) (g6)--(g7) 
(g7)--(g8) (g8)--(g9) (g9)--(g10) (g10)--(g11) (g11)--(g12) (g12)--(g13) 
(g13)--(g14) (g14)--(g15) (g8)--(g16) (g8)--(g18);
	\end{tikzpicture}
	}  
\end{align}
and its Coulomb branch satisfies
\begin{align}
 \MHiggs^{6d}(Q_{n,k}^{\mathrm{D}})\big|_{\{n_i\}} = 
\MCoulomb^{3d}(F_{\{n_i\},k}^{\mathrm{D}}) \; , 
\qquad
 \MCoulomb^{3d}(F_{\{n_i\},k}^{\mathrm{D}}) =  
\MCoulomb^{3d}(F_{\{1^n\},k}^{\mathrm{D}}) \slash \prod_i 
S_{n_i} \; .
\end{align}
Thirdly, the discrete quotient procedure establishes another operation on 
quiver gauge theories solely through their associated Hilbert series. This 
highlights the diverse applicability of the Hilbert series and adds to the 
catalogue of quiver operations such as the ideas of quiver subtraction 
\cite{Cabrera:2018ann} and Kraft-Procesi small instanton transition 
\cite{Hanany:2018uhm}.

In view of other approaches to 3-dimensional $\Ncal=4$ Coulomb branches, like 
the abelianisation method \cite{Bullimore:2015lsa,Bullimore:2016hdc} or the 
attempt to define the Coulomb branch mathematically 
\cite{Nakajima:2015gxa,Nakajima:2015txa,Braverman:2016wma}, it would be 
interesting to understand whether these can reproduce the discrete quotients.

\paragraph{Acknowledgements.}
The authors gratefully acknowledge discussions with Santiago Cabrera, 
Rudolph Kalveks, Anton Zajac, Travis Schedler, and Olaf Krüger. 
We thank the Galileo Galilei Institute for
Theoretical Physics for the hospitality and the INFN for partial support
during the initial stage of this work at the workshop ``Supersymmetric 
Quantum Field Theories in the Non-perturbative Regime'' in 2018.
A.H.\ is supported by STFC Consolidated Grant ST/J0003533/1, and EPSRC 
Programme 
Grant EP/K034456/1.
M.S.\ is supported by Austrian Science Fund (FWF) grant P28590. 
M.S.\ thanks the Faculty of Physics of the University of Vienna for travel 
support via the ``Jungwissenschaftsförderung''.

\appendix
  \section{Background material}
\label{app:all}
\subsection{Cycle index}
\label{app:cycle_index}
The cycle index of a permutation group $\Gamma$ of 
degree $n$ is defined as average of the cycle index monomials of all 
permutations $g\in \Gamma$. Every $g\in \Gamma$ can be decomposed into disjoint 
cycles $c_1 c_2 c_3 \cdots$. Let $j_k(g)$ be the number of cycles in $g$ of 
length $k$, then 
\begin{align}
 Z(\Gamma) = \frac{1}{|\Gamma|} \sum_{g\in \Gamma} \prod_{k=1}^n a_k^{j_{k}(g)} 
\; .
\label{eq:cycle_index_def}
\end{align}
If one considers the symmetric group $S_n$ then cycle index can be cast into a 
recursive relation:
\begin{align}
 Z(S_n) = \frac{1}{n} \sum_{l=1}^n a_l Z(S_{n-l})
 \label{eq:cycle_index_recursive}
\end{align}
where one defines $Z(S_0)=1$. The first recursions yield:
\begin{align}
 Z(S_1)= a_1 \; , \qquad
 Z(S_2)= \frac{1}{2!} (a_2 +a_1^2)  \; , \qquad
 Z(S_3)= \frac{1}{3!} (a_1^3  +3a_1 a_2 +2 a_3 ) \; . 
\end{align}
%
%
\subsection{q-theory}
\label{app:q-theory}
To prove the auxiliary identity
\begin{align}
 Q_l(t) \coloneqq \sum_{j=1}^l \frac{1}{1-t^j} \prod_{a=0}^{j-1} (1-t^{l-a}) 
= l \qquad \forall t \;, 
\label{eq:Q-identity}
\end{align}
one recalls the following definitions from $q$-theory:
\begin{subequations}
\begin{alignat}{2}
 &q\text{-bracket} \qquad&  [k]_q&= \frac{1-q^k}{1-q} \,,\\
 &q\text{-factorial}\qquad &  [k]_q!&= \begin{cases}
   1 &, k=0 \\
  [k]_q \cdot [k-1]_q \cdot \ldots \cdot [1]_q & , k=1,2,\ldots
                                 \end{cases} \; ,\\
 &q\text{-binomial coefficient} \qquad&  \begin{bmatrix}
                                          k \\ j
                                         \end{bmatrix}_q 
&= \frac{[k]_q!}{[j]_q! [k-j]_q!} \,.
\end{alignat}
For the $q$-binomial coefficient exists a $q$-version of the Pascal identities;
for instance
\begin{align}
 \begin{bmatrix} k \\ j\end{bmatrix}_q 
 =
 q^j \begin{bmatrix} k-1 \\ j\end{bmatrix}_q 
 +
 \begin{bmatrix} k-1 \\ j-1\end{bmatrix}_q  \; ,
 \label{eq:q-Pascal}
\end{align}
\end{subequations}
for $j=1,2,\ldots,k-1$. Then, one can rewrite 
\begin{align}
 Q_l(t) &= \sum_{j=1}^l \frac{1}{1-t^j}  
 (1-t^{l})(1-t^{l-1})\cdot \ldots \cdot (1-t^{l-(j-1)}) \notag\\
&=\sum_{j=1}^l \frac{(1-t^{l})(1-t^{l-1})\cdot \ldots \cdot 
(1-t^{l-(j-1)})}{(1-t^{j})(1-t^{j-1})\cdot \ldots \cdot (1-t)} \cdot 
(1-t^{j-1})\cdot \ldots \cdot (1-t) \notag\\
&=\sum_{j=1}^l \begin{bmatrix}l \\ j \end{bmatrix}_t \cdot (1-t)^{j-1} \cdot 
[j-1]_t! 
\; . \label{eq:Q-rewritten}
\end{align}
Having expressed \eqref{eq:Q-identity} as in \eqref{eq:Q-rewritten} has the 
benefit that one can follow an argument of \cite{Quesne:2003}.
The proof proceeds by induction over $l$ employing \eqref{eq:q-Pascal}.
Firstly, the base case is verified easily
\begin{subequations}
\begin{align}
 Q_1(t)&= \begin{bmatrix}1 \\ 1 \end{bmatrix}_t \cdot (1-t)^{0} \cdot 
[0]_t! = 1 \; ,\\
 Q_2(t)&=  \begin{bmatrix}2 \\ 1 \end{bmatrix}_t 
+
 \begin{bmatrix}2 \\ 2 \end{bmatrix}_t \cdot (1-t) \cdot 
[1]_t! = (1+t) + (1-t)=2 \; .
\end{align}
\end{subequations}
Secondly, the inductive step is shown via
\begin{align}
Q_l(t) &=\sum_{j=1}^l \begin{bmatrix}l \\ j \end{bmatrix}_t \cdot (1-t)^{j-1} 
\cdot 
[j-1]_t! \notag\\
&=\sum_{j=1}^{l-1} \begin{bmatrix}l-1 \\ j \end{bmatrix}_t t^j \cdot 
(1-t)^{j-1} \cdot [j-1]_t!
+
\sum_{j=0}^{l-2} \begin{bmatrix}l-1 \\ j \end{bmatrix}_t \cdot (1-t)^{j} 
\cdot [j]_t! \notag\\
&=\sum_{j=1}^{l-1} \begin{bmatrix}l-1 \\ j \end{bmatrix}_t t^j \cdot 
(1-t)^{j-1} \cdot [j-1]_t! +1
+
\sum_{j=1}^{l-1} \begin{bmatrix}l-1 \\ j \end{bmatrix}_t \cdot (1-t)^{j-1} \cdot
(1-t^j) \cdot [j-1]_t! \notag\\
&=\sum_{j=1}^{l-1} \begin{bmatrix}l-1 \\ j \end{bmatrix}_t \cdot 
(1-t)^{j-1} \cdot [j-1]_t! +1 \notag \\
&= Q_{l-1}(t) +1 = (l-1)+1
\end{align}
where the induction hypothesis has been used in the last step. 

 \bibliographystyle{JHEP}     %
 {\footnotesize{\bibliography{references}}}

\end{document}